\documentclass{article}

\usepackage{arxiv}

\usepackage[utf8]{inputenc} 
\usepackage[T1]{fontenc}    
\usepackage{hyperref}       
\usepackage{url}            
\usepackage{booktabs}       
\usepackage{amsfonts}       
\usepackage{nicefrac}       
\usepackage{microtype}      
\usepackage{lipsum}		
\usepackage{graphicx}
\usepackage{natbib}
\usepackage{doi}
\usepackage{amsmath}
\usepackage[title]{appendix}

\title{Variational Inference at Glacier Scale}

\author{Douglas J. Brinkerhoff \\
	Department of Computer Science\\
	University of Montana\\
	Missoula, MT 59812 \\
	\texttt{douglas1.brinkerhoff@umontana.edu} \\
}

\renewcommand{\shorttitle}{VI at Glacier Scale}

\hypersetup{
pdftitle={Variational Inference at Glacier Scale},
pdfsubject={physics.comp-ph},
pdfauthor={Douglas J. Brinkerhoff},
pdfkeywords={Variational Inference, Ice Sheet Modelling, Bayesian Inference},
}

\begin{document}
\maketitle

\begin{abstract}
We characterize the complete joint posterior distribution over spatially-varying basal traction and and ice softness parameters of an ice sheet model from observations of surface speed by using stochastic variational inference combined with natural gradient descent to find an approximating variational distribution.  By placing a Gaussian process prior over the parameters and casting the problem in terms of eigenfunctions of a kernel, we gain substantial control over prior assumptions on parameter smoothness and length scale, while also rendering the inference tractable.  In a synthetic example, we find that this method recovers known parameters and accounts for mutual indeterminacy, both of which can influence observed surface speed.  In an application to Helheim Glacier in Southeast Greenland, we show that our method scales to glacier-sized problems.  We find that posterior uncertainty in regions of slow flow is high regardless of the choice of observational noise model.        
\end{abstract}


\section{Introduction}
Glaciers and ice sheets are undergoing mass loss as a result of warming-induced perturbations to precipitation and melt rates and adjustments in flow speed \citep{shepherd2018antarctica,shepherd2020greenland,gardner2013reconciled}.  As a means of understanding and predicting the quality and quantity of such changes, physics-based models of ice flow have emerged as a standard tool.  Ice is typically modelled as a slow fluid that is both nonlinear in viscosity and thermo-mechanically sensitive.  These non-linearities result in substantive computational challenges on their own, but are exacerbated by two even more critical epistemic challenges.  First, the majority of glacier motion is the result of motion that occurs at the basal boundary, either through explicit slip or through deformation of an underlying till (or perhaps both), and a validated constitutive relationship describing this process is not known.  Such a relationship may depend sensitively on a complex hydrologic system beneath the ice \citep[e.g][]{iken1986combined,bindschadler1983importance, fowler1979mathematical} and spatially variable material properties \citep[e.g][]{zoet2020slip,truffer2001implications,macayeal1989large}.  Second, effects occurring across lengths and time scales of millimeters and years to kilometers and millennia exert strong effects on the rheology of the ice.  For example, variations in the conversion of snow to ice \citep{gundestrup1984bore}, the presence of englacial water \citep{duval1977role}, and accumulated damage \citep{duddu2012} (all in addition to the rheologic effects of thermal variability subject to its own highly uncertain forcing \citep{harrington2015}) may all lead to non-negligible changes in ice rheology that are difficult to predict.  

Both of these knowledge gaps are difficult to close at scale.  While much work has been done in trying to observe glacier sliding and viscous flow \emph{in situ} as a means (either directly or indirectly) of improving the treatment of these processes in flow models, the realities of the glacial environment enforce spatial and temporal sparsity on such observations; there is no plausible analogue to the weather balloon or moorings that strongly constrain earth's other large-scale surface fluids.  As such, inverse modelling, here defined as the process of inferring unknown quantities (such as basal traction and ice rheologic prefactor) from those that are more easily observed (such as ice surface velocities derived by tracking recurrent features between repeated satellite images), has become a critical component in constraining ice sheet flow towards making more realistic predictions of change over a wide range of spatial and temporal scales.

Unfortunately, such inverse problems are ill-posed in the sense that a particular realization of random errors applied to observations (e.g. noisy surface velocities from satellite observations), can lead to a substantially different solution for the inferred field than a different realization.  Furthermore, qualitatively distinct unknown factors can compensate for one another, leading to a fundamental indeterminacy (e.g. the same observed surface velocity could be produced by a slippery bed and stiff ice or by a sticky bed and fluid ice).  While the imposition of constraints on smoothness and magnitude of such factors (so-called regularization) can act to \emph{reduce} this non-uniqueness, it cannot eliminate the fact that our state of knowledge about an unknown quantity conditioned upon imperfect and indirect measurements are more appropriately quantified as a probability distribution.  This latter conclusion is the fundamental motivation behind Bayesian inference; \textbf{to approximate this probability distribution in a manner that is accurate, complete, and computationally tractable is the primary goal of this paper}.  

In the following sections we formally define the statement of the Bayesian inference problem, and specify the prior and likelihood distributions that constitute its parts.  We use Gaussian Process priors \citep{williams2006gaussian}, which are flexible probability distribution over spatial functions that nonetheless come equipped with easily understood assumptions about critical quantities like length scale and smoothness.  Recognizing that our problem is too large for their standard application, we instead form an approximation in terms of Fourier-like spectra that leads to a significant reduction in the problem's computational complexity \citep{solin2020hilbert}.  Next, we introduce the likelihood function that relates the traction coefficient and rheologic prefactor to observed surface speeds through the highly non-linear ice flow equations.  This relationship (along with potential non-normality of observational errors) means that the inverse problem has no analytical solution.  We instead develop a stochastic variational inference method \citep{blei2017variational,kucukelbir2016automatic,blundell2015weight,hoffman2013stochastic}, in which we hypothesize the family of an approximating ansatz distribution (with its own free parameters), and adjust it via natural gradient descent such that the approximating distribution minimizes a metric of divergence from the true posterior.  We apply our method to a set of simple analytical problems \citep[the ISMIP-HOM experiments][]{pattyn2008benchmark} before turning to the inference of the traction coefficient and rheological prefactor at Helheim Glacier in Southeast Greenland.    

\section{Bayesian Inference of Basal Shear Stress}

\begin{figure}
    \centering
    \includegraphics[width=1.0\linewidth]{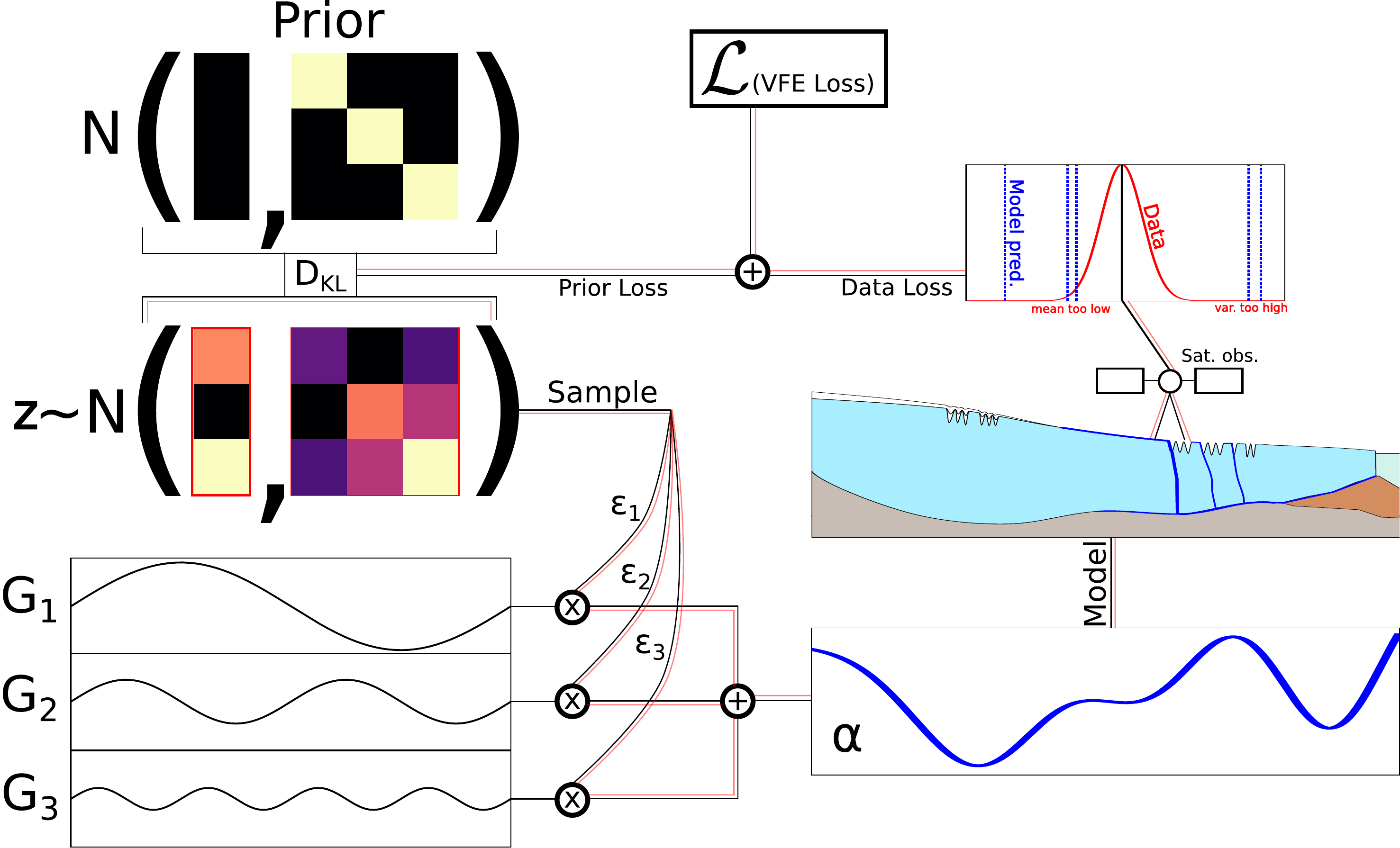}
    \caption{An overview of the workflow for variational inference presented in this work, with black lines representing the forward flow of information, and red lines the inverse flow.  The variable to be inferred is $\mathbf{z}$, which is parameterized as a normal distribution.  Values of $\mathbf{z}$ act as coefficients to a spectral representation of the spatially varying parameters $\alpha$ and $\tau$ (latter not shown), which in turn are mapped to a surface velocity through an ice sheet model.  The ice sheet model predictions are compared to observations to form a data loss, which is minimized alongside a prior loss, which tries to keep the $\mathbf{z}$ a unit normal distribution.}
    \label{fig:overview}
\end{figure}
We seek the probability distribution
\begin{equation}
    \underbrace{P(\beta^2, A | \mathbf{u}_{obs})}_{\mathrm{Posterior}} \propto \underbrace{P(\mathbf{u}_{obs} | \beta^2, A)}_{\mathrm{Likelihood}} \underbrace{P(\beta^2, A)}_{\mathrm{Prior}},
\end{equation}
where $\beta^2(\mathbf{x})$ is the basal traction coefficient, $A(\mathbf{x})$ is the ice viscosity prefactor, and $\mathbf{u}_{obs}$ is the collection of observed surface velocities.  We note that both parameters are functions of the spatial coordinates $\mathbf{x}$, but we henceforth omit this explicit dependence unless necessary to reduce clutter.  The left-hand side of this relation is the posterior distribution, which represents joint knowledge of traction and viscosity after the consideration of data.  Bayes' theorem gives the right hand side, where the first term is the likelihood (the probability of having observed the data assuming a given value of $\beta^2$ and $A$) and the second term is the prior (our knowledge $\beta^2$ and $A$ prior to having considered $\mathbf{u}_{obs}$).  

\subsection{Reparameterization}
Both $\beta^2$ and $A$ are positive quantities that can vary over several orders of magnitude in typical glaciological scenarios.  As such, we introduce a logarithmic reparameterization of both variables:
\begin{align}
    \beta^2 &= \beta^2_0 \mathrm{exp}(\alpha) \\
    A &= A_0 \mathrm{exp}(\tau),
\end{align}
where $\beta_0,A_0$ are reference values (e.g. at the pressure melting point).  We thus rewrite Bayes theorem as 
\begin{equation}
    P(\alpha, \tau | \mathbf{u}_{obs}) \propto P(\mathbf{u}_{obs} | \alpha, \tau) P(\alpha, \tau).
    \label{eq:bayes}
\end{equation}
This reparameterization renders both $\alpha$ and $\tau$ unconstrained and the exponential variability characteristic to traction and viscosity is now rendered approximately linear in these new parameters, which simplifies the specification of the prior distribution.  
\subsection{The Prior}
To use Eq.~\ref{eq:bayes}, we must explicitly specify distributions for each of the terms on the right hand side.  We begin with the prior distribution $P(\alpha,\tau)$.  As $\alpha$ and $\tau$ are functions, we require that each be modeled as a distribution over functions.  One flexible choice is to model $\alpha$ and $\tau$ as Gaussian Processes (GPs) \citep{williams2006gaussian}.  A GP is a random function in which the function values at any finite collection of evaluation points are distributed as a multivariate normal distribution
\begin{equation}
    P(f(\mathbf{X})) = \mathcal{N}(\mu(\mathbf{X}), K(\mathbf{X},\mathbf{X}')),
\end{equation}
where $\mathbf{X} \in \Omega$ is a finite vector of spatial coordinates (for example, the nodal coordinates of a finite element mesh) inside a spatial domain $\Omega$, $\mu(\mathbf{x})$ is a mean function, and $K(\mathbf{X},\mathbf{X}')$ is the covariance matrix produced by pairwise evaluation of a covariance function (or kernel) $k(\mathbf{x},\mathbf{x}')$.  The choice of kernel imparts specific properties on the resulting functions.  In this work we use the squared exponential kernel
\begin{equation}
    k_{SE}(\mathbf{x},\mathbf{x}') =  \sigma^2 \mathrm{exp}\left(\frac{|\mathbf{x} - \mathbf{x}'|^2}{2 \ell^2} \right),
\end{equation}
where $\ell$ is the length scale (which governs the distance beyond which function values are uncorrelated), and $\sigma^2$ is the variance.  The squared exponential kernel is infinitely differentiable, and thus produces functions that are smooth.  For points that are close in space, the kernel produces strongly correlated function values, while distant points are uncorrelated.

We assume \emph{a priori} that $\alpha$ and $\tau$ are independent and place a GP prior on each.  Both have zero mean and independent kernel hyperparameters  (to be defined for individual experiments).  In particular, we expect that basal shear stress should vary over substantially shorter length scales than ice viscosity, and these independent length scales allow us to encode this.  We write the joint prior as  
\begin{equation}
    P(\alpha, \tau) = \mathcal{N}\left( \begin{bmatrix} \mathbf{0} \\ \mathbf{0} \end{bmatrix},\begin{bmatrix} K_\alpha & \mathbf{0} \\ \mathbf{0} & K_{\tau} \end{bmatrix} \right),
    \label{eq:joint_prior}
\end{equation}
where $\mathbf{0}$ implies an appropriately sized matrix or vector of zeros, and $K_\alpha$ and $K_\tau$ are the covariance matrices for their respective functions.

\subsection{Low rank representation}
Unfortunately, Eq.~\ref{eq:joint_prior} cannot immediately be applied to large problems due to the need to form and manipulate the large and dense covariance matrices $K_{\alpha}, K_{\tau} \in \mathbb{S}_{++}^{n}$ (the set of positive definite matrices of size $n$) where $n$ is the number of index points in $\mathbf{X}$.  However, such matrices are often (in an approximate sense) low rank, which is to say that the information contained therein can be represented more compactly.  Because covariance matrices are positive definite, they can be decomposed as 
\begin{equation}
    K_\diamond = V_\diamond \mathrm{diag}(\mathbf{\lambda}_\diamond) V_\diamond^T,
    \label{eq:eigendecomposition}
\end{equation}
where the $\diamond$ subscript implies either $\alpha$ or $\tau$, $\lambda_\diamond \in \mathbb{R}_+^n$ is a vector containing the (real and positive) eigenvalues of $K_\diamond$ ordered from largest to smallest, and $V_\diamond \in \mathbb{R}^{n\times n}$ the associated (real and orthonormal) eigenvectors.  For a low rank matrix, the diagonal entries of $\lambda_\diamond$ rapidly decay towards zero.  Truncating these eigenvalues at index $m_\diamond$ such that $\sqrt{\frac{\lambda_j}{\lambda_0}} < r$ for $j> m_\diamond$ and $r$ a tolerance yields the reduced eigendecomposition
\begin{equation}
    K_\diamond \approx \hat{V}_\diamond \mathrm{diag}(\hat{\mathbf{\lambda}})_\diamond \hat{V}_\diamond^T,
\end{equation}
where $\hat{\mathbf{\lambda}}_\diamond \in \mathbb{R}^{m}$ and $\hat{V}_\diamond \in \mathbb{R}^{n\times m}$.  Writing 
\begin{equation}
    G_\diamond = \hat{V}_\diamond \mathrm{diag}\sqrt{\mathbf{\hat{\lambda}}}_\diamond,
\end{equation}
we have the generative model
\begin{align}
    P(\mathbf{z}_\alpha) &= \mathcal{N}(\mathbf{0},\mathbf{I}) \\
    \alpha &= G_\alpha \mathbf{z}_\alpha,
\end{align}
\begin{align}
    P(\mathbf{z}_\tau) &= \mathcal{N}(\mathbf{0},\mathbf{I}) \\
    \tau &= G_\tau \mathbf{z}_\tau,
\end{align}
where $\mathbf{z}_\diamond$ is a latent variable of dimension $m_\diamond$ with zero mean and identity covariance matrix.  The matrix $G_\diamond$ is a feature map in that it (approximately) lifts the index points $\mathbf{X}$ into the reproducing kernel Hilbert Space associated with the kernel $k_\diamond(\mathbf{x},\mathbf{x})$ .  For the sake of intuition, we can think of latent variables $\mathbf{z}_\cdot$ as the coefficients of a Fourier series scaled by a function mapping frequency to amplitude.  Indeed, this analogy is exact for 1D problems with periodic boundary conditions.  

In this alternative representation, we seek to infer the coefficients $\mathbf{z} = [\mathbf{z}_\alpha,\mathbf{z}_\tau]^T$, with Bayes theorem giving 
\begin{equation}
    P(\mathbf{z} | \mathbf{u}_{obs}) \propto P(\mathbf{u}_{obs} | \mathbf{z} ) P(\mathbf{z}),
    \label{eq:bayes_reparameterized}
\end{equation}
with the explicit prior
\begin{equation}
    P(\mathbf{z}) = \mathcal{N}\left(\begin{bmatrix} \mathbf{0} \\ \mathbf{0}\end{bmatrix}, \begin{bmatrix} \mathbf{I} & \mathbf{0} \\ \mathbf{0} & \mathbf{I} \end{bmatrix}\right).
\end{equation}

We note that computing the $\mathcal{O}(n^2 m_\diamond)$ eigendecomposition Eq.~\ref{eq:eigendecomposition} is intractable for large problems (i.e. $n>10^4$).  To this end, we construct an approximate eigendecomposition of $K_\diamond$ using the Hilbert space method of \citet{solin2020hilbert}, which constructs the columns of $G_\diamond$ as scaled eigenfunctions of the Laplacian operator.  A detailed description of this construct can be found in Appendix 1.

\subsection{Likelihood Model}
As the second ingredient in the posterior, we turn to defining the likelihood model $P(\mathbf{u}_{obs} | \mathbf{z})$, which quantifies the probability of measuring the data given a specified value for $\mathbf{z}$.  It is not well understood what distribution appropriately models observational uncertainty in surface velocities derived from synthetic aperture radar, and so we adopt the following explicit assumptions.  We hypothesize two different models of pointwise observational noise depending on the experiment.  First, a Gaussian noise model 
\begin{equation}
    P(\mathbf{u}_{obs}(\mathbf{x}) | \mathbf{z}) = \mathcal{N}(\mathbf{u}_{obs}(\mathbf{x}); \mathbf{u}(\mathbf{x};\mathbf{z}), \sigma^2_{obs}),
    \label{eq:normal_likelihood}
\end{equation}
where $\sigma^2_{obs}$ is the observational variance and $\mathbf{u}(\mathbf{x}; \mathbf{z})$ is a prediction function that maps from spectral coefficients $\mathbf{z}$ to predicted surface velocities.  As described above, we obtain explicit spatial fields of the traction coefficient and rheologic prefactor by applying feature maps $G_\alpha$ and $G_\tau$ and exponentiating.  Given these parameters, we then employ an ice flow model that numerically solves a set of partial differential equations via the finite element method to make predictions of the velocity given those parameters.  A detailed description of this model is given in Appendix 2, but here we have written it as a function $\mathbf{u}(\mathbf{x};\mathbf{z})$ that takes as input a spatial coordinate and parameter vector and produces a surface velocity.  

Alternatively, we can employ the log-normal noise model
\begin{equation}
    P(\ln |\mathbf{u}_{obs}(\mathbf{x})|\; |\mathbf{z}) = \mathcal{N}(\ln |\mathbf{u}_{obs}(\mathbf{x})|;\; \ln |\mathbf{u}(\mathbf{x};\mathbf{z})|, \sigma^2_{r,obs}),
\end{equation}
where $\sigma^2_{r,obs}$ is the variance of the logarithm of the velocity, which can be interpreted as relative error in the sense that $\sigma_{r,obs}=0.1$ corresponds to a standard standard deviation in speed of $\sim 10\%$.  These two models can be viewed as modelling the observational uncertainty as absolute (the former) or relative (the latter).  We note that the log-normal distribution is substantially easier to optimize when velocities vary over multiple orders of magnitude (as in many real glaciers).  There are many other candidate distributions that could be plausibly be employed as models of observational noise, such as the $t$ \citep[which was discussed as model of velocity noise in][]{gopalan2021bayesian} or Laplace distributions; a more careful and systematic investigation of this choice is warranted in future studies.  

The likelihood describes a single observation.  For multiple discrete observations, we assume independence, and the full likelihood is given by a simple product over observations.  However, in practice, surface velocity observations are typically given as observations interpolated to a grid, where the grid spacing is not necessarily equivalent to the `true' number of independent observations.  Identification of observation points with the nodes of a computational mesh (through interpolation) leads to grid dependence.  In this work, we instead formulate the likelihood as an integral
\begin{equation}
    \ln P(\mathbf{u}_{obs} | \mathbf{z}) = \int_\Omega \ln P(\mathbf{u}_{obs}(\mathbf{x}) | \mathbf{z}) \mathrm{d} \mathbf{x},
\end{equation}
where we have introduced the observation density parameter $\rho$, which is a hyperparameter describing the number of independent observations per unit area.  This formulation is in good conceptual agreement with the integral formulation of cost functions that normally appear in PDE-constrained optimization problems in glaciology, and also helps to ensure that interpolation and post-processing of gridded observations does not unduly influence the solution of the inverse problem.  Nonetheless, more effort is needed to find a justified model of the spatial dependence of velocity noise.

\section{Variational Inference}
Because of the non-linearity of the ice sheet model and the non-normality of the likelihood function, the posterior distribution cannot be evaluated in closed form, and as such must be approximated.  The approach that we use for this approximation here is variational inference \citep{blei2017variational}.  In variational inference, we assume that the posterior distribution belongs to a family of tractable distributions with learnable variational parameters.  We then minimize a divergence metric between that distribution and the true posterior with respect to these variational parameters.  This turns Bayesian inference into an optimization problem, which is both less computationally expensive than MCMC methods \citep{brooks2011handbook} but also much more expressive than simpler approximations that rely on, for example, the posterior's local curvature as in the Laplace approximation.  

We begin by defining the the variational distribution over the random variable $\mathbf{z}$ that we will fit to the posterior.  Because the prior distribution is normally distributed (specifically, a GP), we choose to use the normal distribution
\begin{equation}
    q_{\xi}(\mathbf{z}) = \mathcal{N}(\mathbf{z};\mu, S),
\end{equation}
where $\mu$ is a mean vector of size $m = m_{\alpha} + m_{\tau}$ and $S$ is a full-rank, symmetric, positive definite covariance matrix.  We use $\xi=[\mu,\mathrm{vech}(\log S)]^T$ to denote the $m + m(m-1)/2$ vector of variational parameters, where $\mathrm{vech}$ is the half-vectorization operator \citep{magnus2019matrix}.  We choose to parameterize the covariance matrix via its matrix logarithm $\ln S$ to ensure that the resulting covariance (the matrix exponential of $\ln{S}$) is positive definite; if $\Theta$ and $\gamma$ are the eigenvectors and eigenvalues of $\log S$, then the covariance matrix $S$ is 
\begin{equation}
    S = \Theta \,\mathrm{diag}(\mathrm{e}^{\gamma}) \,\Theta^T.
\end{equation}
We emphasize that although the prior distribution over $\mathbf{z}$ is diagonal (meaning that we assume no prior correlation between the modes of $\alpha$ or $\tau$ with themselves or one another), the variational distribution $q_\xi(\mathbf{z})$ is not restricted as such: this is important because both parameters can affect surface velocities in similar ways, and will be unidentifiable.  We also note that while the variational distribution is defined over coefficients of feature space, we can (nominally) recover the posterior distribution over $[\alpha,\tau]^T$ as 
\begin{equation}
    P\left(\begin{bmatrix} \alpha \\ \tau \end{bmatrix}\bigg|\mathbf{u}_{obs}\right) = \mathcal{N}(\mathbf{z}; G\mu, G S G^T),
\end{equation}
although the covariance (which is $2n\times 2n$) may be too large to form explicitly, although we can easily investigate marginal distributions by only computing the diagonal. 

The statement of the variational distribution is simply an assumption about the form of the posterior; to make it useful, we must find appropriate values of the variational parameters.  To accomplish this goal we minimize the Kullback-Leibler divergence from the posterior $P(\mathbf{z}|\mathbf{u}_{obs})$ to $q_\xi(\mathbf{z})$, which is given by
\begin{equation}
    D_{KL}\bigg(q_\xi(\mathbf{z})||P(\mathbf{z}|\mathbf{u}_{obs})\bigg) = \int q_\xi(\mathbf{z}) \ln{\frac{P(\mathbf{z}|\mathbf{u}_{obs})}{q_\xi(\mathbf{z})}} \mathrm{d} \mathbf{z}.
\end{equation}
The KL divergence measures the amount of information lost by approximating the true posterior distribution with the variational posterior, and is clearly minimized when the two are equal.  Because the true posterior is not known, the KL divergence cannot be evaluated.  However, using Bayes theorem on the posterior we derive an equivalent expression 
\begin{equation}
    D_{KL}\bigg(q_\xi(\mathbf{z})||P(\mathbf{z}|\mathbf{u}_{obs})\bigg) = \int q_\xi(\mathbf{z}) \bigg(\ln P(\mathbf{u}_{obs}|\mathbf{z}) + \ln P(\mathbf{z}) - \ln q_\xi(\mathbf{z}) - \ln P(\mathbf{u}_{obs})\bigg) \mathrm{d}\mathbf{z}.  
\end{equation}
$\ln P(\mathbf{u}_{obs})$ (sometimes referred to as the log-evidence) is independent of the variational parameters $\xi$ and thus can be neglected.  The resulting function is the \emph{variational free energy}
\begin{equation}
    \mathcal{L}\bigg(q_\xi(\mathbf{z}),P(\mathbf{z}|\mathbf{u}_{obs})\bigg) = \underbrace{\int q_\xi(\mathbf{z}) \ln P(\mathbf{u}_{obs}|\mathbf{z})\mathrm{d}\mathbf{z}}_{\text{Data Loss}} + \underbrace{D_{KL}\bigg(q_\xi(\mathbf{z}) || P(\mathbf{z})\bigg)}_{\text{Prior Loss}}
\end{equation}
Minimizing the variational free energy thus represents a trade-off between minimizing the expected observational misfit with respect to the parameters (the first term) while maintaining fidelity to the prior distribution (the second term).  The prior distribution over $\mathbf{z}$ is by construction a unit normal distribution, which admits a simple analytical form for the KL divergence from the variational distribution to the prior
\begin{equation}
    D_{KL}\bigg(q_\xi(\mathbf{z}) || P(\mathbf{z}\bigg) = \frac{1}{2}\left(\mu^T \mu + \mathrm{tr}(S) - m - \log |S|\right).
\end{equation}
The integral involving the likelihood term is still intractable.  However, we can approximate the expectation via the simple unbiased Monte Carlo integration \citep{kucukelbir2016automatic,blundell2015weight}
\begin{align}
    \int q_\xi(\mathbf{z}) \ln P(\mathbf{u}_{obs}|\mathbf{z})\mathrm{d}\mathbf{z} &\approx \frac{1}{k} \sum_{i=1}^k \ln P(\mathbf{u}_{obs}|\mathbf{z}_i) \\
    \mathbf{z}_i &\sim q_\xi(\mathbf{z}).
\end{align}
Because we will use an iterative method to minimize the VFE, we use $k=1$, i.e. we approximate this integral with a single random sample every time we evaluate the VFE.  

\subsection{Natural gradient descent}
To minimize the VFE, we employ the gradient descent method
\begin{align}
  \xi_{t+1} = \xi_t - \eta p_t,
\end{align}
with $\eta$ a step size and $p$ a search direction, repeating the process until a norm of parameter changes falls below a threshold.  So-called `vanilla' gradient descent defines
\begin{equation}
    p_t = \nabla_\xi \mathcal{L}_t,
\end{equation}
or the Euclidean gradient of the variational free energy with respect to the mean and log-covariance of the variational distribution.  However, the Euclidean metric is a poor measure of distance between probability distributions.  For example, consider two unit normal distributions with means $\mu_0=0$ and $\mu_1=1$, and each with a variance much less than 1.  Clearly, these distributions are quite different.  Now consider two unit normal distributions with the same means, but variance much greater than 1.  These two distributions are quite similar: the qualitative difference between means depends on the variance, but this intuition is not captured with vanilla gradient descent.  Natural gradient descent captures this distinction by defining the descent direction as the gradient using the KL divergence between distributions at iteration $t$ and $t+1$ as a metric.  In practical terms, this means that the search direction is preconditioned with the inverse of the Fisher information matrix $F$
\begin{equation}
    p_t = F_t^{-1} \nabla_\xi \mathcal{L}_t
\end{equation}
where $F = \mathbb{E}_{q_\xi(\mathbf{z})}[\nabla^2 q_\xi(\mathbf{z})]$.
Unfortunately, even for the reduced rank parameterization considered here, it is not possible to store this matrix explicitly, much less invert it.  However, \citet{salimbeni2018natural} show that the natural gradient update can be efficiently calculated for an arbitrary parameterization as
\begin{equation}
    p = \frac{\partial \xi}{\partial \theta} \nabla_\eta \mathcal{L},
    \label{eq:salim}
\end{equation}
where $\eta$ is the expectation parameterization of the variational distribution
\begin{align}
    \eta = \begin{bmatrix} \mu\\ \mathrm{vech}(S + \mathbf{\mu}\mathbf{\mu}^T) \end{bmatrix}
\end{align}
and $\theta$ is the natural parameterization
\begin{align}
    \theta = \begin{bmatrix} S^{-1} \mu \\ \mathrm{vech}\left(-\frac{1}{2} S^{-1}\right)\end{bmatrix},
\end{align}
with $\nabla_\eta$ the gradient with respect to the expectation parameterization.  As a matter of practical computation we compute the right hand side by first transforming our native parameterization $\xi$ to the expectation parameterization $\eta$.  Saving the computational graph to allow automatic differentiation, we then convert $\eta$ to a parameterization in terms of the mean $\mu$ and a Cholesky factor $L$ such that $L L^T = S$.  The Cholesky parameterization is particularly convenient because it allows us to sample from $q_\xi(\mathbf{z})$ easily using the reparameterization trick
\begin{equation}
    \mathbf{z}_i = \mu + L \epsilon,
\end{equation}
where $\epsilon$ is unit Gaussian noise.  Note that this eliminates the stochasticity from the variational parameters and ensures that they have a well-defined gradient.  Having sampled from $q_\xi(\mathbf{z})$ in a differentiable way, we can now evaluate the VFE and reverse-mode automatic differentiation to compute $\nabla_\eta \mathcal{L}$.  We note that back-propagation through the ice sheet model is performed by automatic symbolic derivation of its adjoint \citep{macayeal1993tutorial}.

On first glance, it appears that we must still form $\frac{\partial \xi}{\partial \theta}$ with $\mathcal{O}(m^4)$ entries.  However, this term is the Jacobian of $\xi(\theta)$ (where we use this overloaded notation to indicate that $\xi$ is a function of $\theta$) and the right hand side of Eq.~\ref{eq:salim} is a Jacobian-vector product.  We thus utilize the fact that Jacobian-vector products can be approximated with the finite difference method \citep{knoll2004jacobian}
\begin{equation}
    p_t \approx \frac{\xi(\theta + \delta \nabla_\eta \mathcal{L}) - \xi(\theta)}{\delta},
    \label{eq:fd}
\end{equation}
with $\delta$ a small constant, here taken as $\delta=10^{-8}$.  The finite difference approximation given in Eq.~\ref{eq:fd} allows relatively efficient computation of the natural gradient search direction, but computing this search direction is still expensive compared to vanilla gradient descent due to the need to perform several $\mathcal{O}(m^3)$ operations in the course of the requisite parameter transformations.  Nonetheless, because of the strong correlations between parameters, particularly the entries of the covariance matrix, this higher order method is essential in achieving convergence.    

\subsection{BFGS initialization}
Before running the stochastic optimization technique described above, we initialize the variational mean $\mu$ from a MAP point by solving a deterministic minimization problem (interestingly, by setting the noise vector $\epsilon$ to zero, we recover the classic non-stochastic optimization problem with Tikhonov regularization typical of glaciology problems, although still defined in terms of coefficients in Fourier-like space).  While not strictly necessary, this does speed the optimization procedure since the stochastic optimization has, by definition, noisier gradients, and does not allow for globalization methods like line search.  We use the standard quasi-Newton algorithm BFGS \citep{nocedal2006numerical}, which returns an optimal value for $\mu$ as well as a low rank approximation of the inverse Hessian, which can be identified as the covariance approximation.  While this latter approximation is not particularly good (we find that it greatly underestimates the posterior covariance), we use it as an initial guess for the variational distribution's covariance $S$.  

\section{Idealized Experiments: ISMIP-HOM}
As an initial illustrative example, we use the setup of the ISMIP-HOM experiments \citep{pattyn2008benchmark} (in particular the flow-line variants B and D) as a test bed for understanding the qualitative characteristics of the method, and to establish a correspondence between the posterior distributions produced by this method and a known true value in the absence of model mis-specification.  While ostensibly similar, this similarity is superficial: they are quite different with respect to the dynamics that they test: Experiment B features a high and constant traction, with substantial internal deformation, while Experiment D has traction that varies strongly in space over several orders of magnitude, with very little shearing.  Both experiments exhibit the effects of longitudinal stress, which act as a low pass filter on variations in traction coefficient, but which also depend sensitively on the viscosity.  Critically, the chosen frequency of sinusoidal variations in topography or friction scales the relative importance of longitudinal stresses.  By varying this frequency, we gain insight into the characteristic spatial scales that are informed by surface observations.  
\subsection{ISMIP-HOM B}

\begin{figure}
    \centering
    \includegraphics{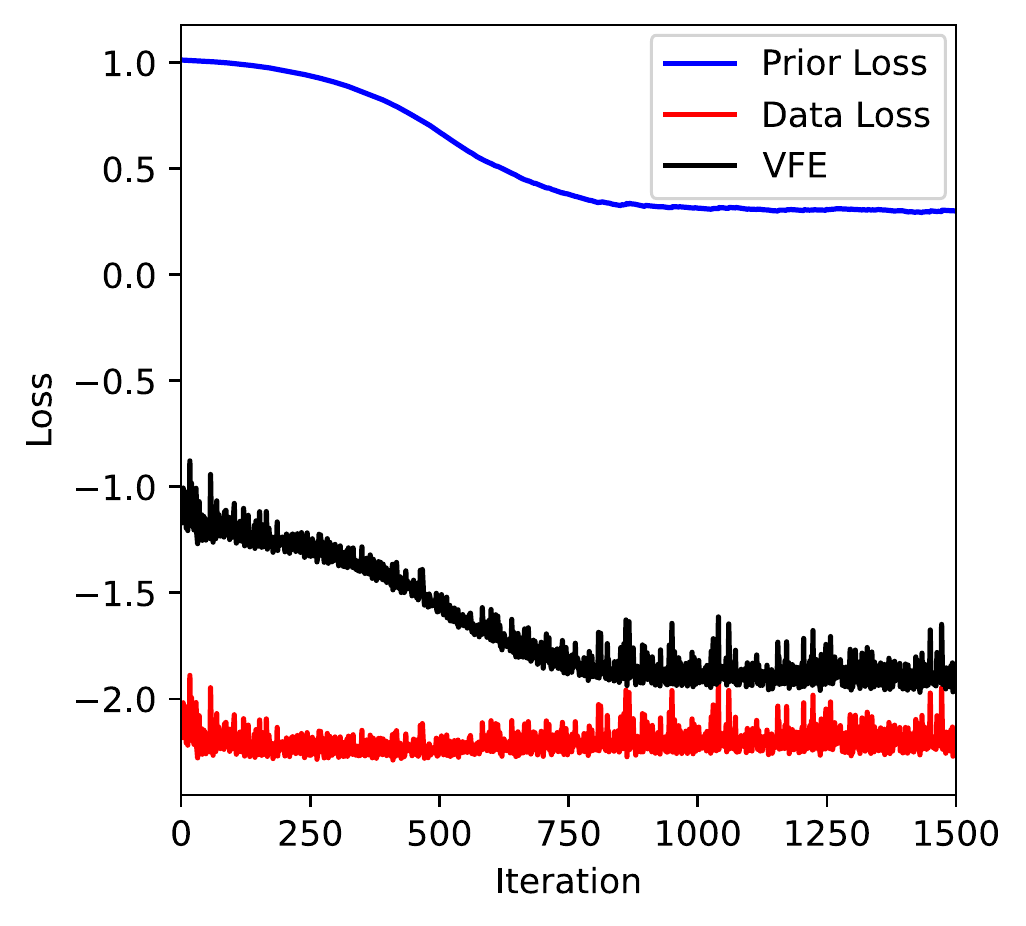}
    \caption{Components of the variational free energy as a function of natural gradient descent iteration for ISMIP-HOM experiment B with $f=8$.  The likelihood component of the loss stays relatively constant, indicating that the initial guess for the mean of the variational distribution computed by BFGS is close to optimal.  The decrease in the prior loss (and therefore the total loss) is caused by an inflation in the posterior approximation's covariance matrix, which is encouraged to be similar to the prior.  This is also seen in the likelihood loss becoming more noisy for later iterations, which reflects noisier random samples associated with a larger uncertainty}
    \label{fig:convergence}
\end{figure}

\begin{figure}
    \centering
    \includegraphics{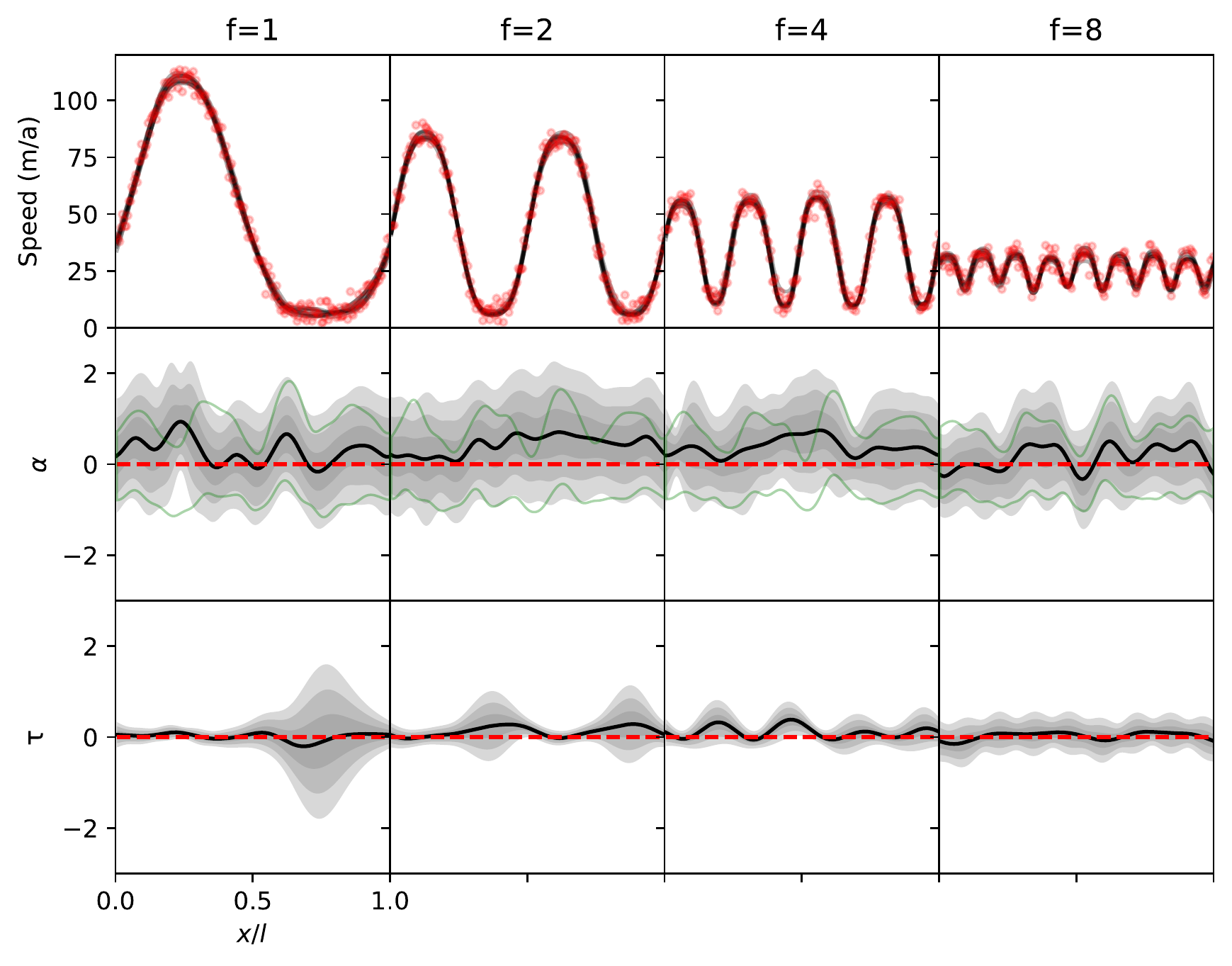}
    \caption{The posterior distribution for $f\in\{1,2,4,8\}$ for ISMIP-HOM B.  In the top row, red dots indicate observed velocity values, while overlain black lines are random samples from the posterior predictive distribution over velocity.  In the bottom two rows, the dashed red line represents the `true' parameter value, while the gray envelopes represent the 1$\sigma$, 2$\sigma$, and 3$\sigma$ marginal credibility intervals, with the black line the posterior mean.  The green lines represent the 3$\sigma$ marginal credibility interval when $\tau$ is assumed known.}
    \label{fig:ismip_b_posterior}
\end{figure}
We define ISMIP-HOM B over the domain $\mathbf{x} = x \in [0,L]$ with periodic boundary conditions, where we take $L=80$km.  The surface elevation is given by 
\begin{equation}
    z_s = -x \tan \alpha,
\end{equation}
where $\alpha=0.5^{\circ}$, and the bed elevation as 
\begin{equation}
    z_b = z_s - 1000 + 500\sin\left(\frac{2\pi f \mathbf{x}}{L}\right),
\end{equation}
where $f\in[1,2,4,8]$ corresponds to the frequency of basal topography undulations.  We note that this definition of the ISMIP-HOM experiment is somewhat different to the typical definition, in that we keep the domain size fixed and adjust the frequency of the bed topography, rather than adjust the domain length to correspond to one sinusoid.  We do this in order to more easily maintain consistency with respect to observation density and prior length scale.

We generate pseudo-observations by running the forward model on known parameters $\beta^2=10^4$ (which approximately corresponds to a no-slip condition) and $A=10^{-16}$, then corrupting the nodal solutions with noise
\begin{equation}
\mathbf{u}_{obs} = \mathbf{u}(\beta^2, A) + \epsilon
\end{equation}
\begin{equation}
    \epsilon \sim \mathcal{N}(\mathbf{0},\sigma^2_{obs} \mathcal{I}).
\end{equation}

We use the normally distributed likelihood model Eq.~\ref{eq:normal_likelihood}, assume knowledge of the true data-generating process, and model it correctly (a big and questionable assumption in the real world, but one which allows us to more easily observe the method's behavior).  As a prior on $\alpha$ we take $\sigma^2=1$ and length scales $l_\alpha=\frac{L}{20}=8$km.  As a prior on $\tau$ we take $\sigma^2_\tau=1$ and $l_{\tau}=\frac{L}{10}=16$km.  We take $\beta_0=10^4$ and $A_0=10^{-16}$.  We assume a data density of $\rho=1$ observation km$^{-1}$.  

We use the Hilbert space method to find a reduced rank basis set.  However, due to the simple geometry and periodic boundary conditions, this basis set takes a particularly simple form, namely a truncated Fourier series.  We find that $m_\alpha=33$ and $m_{\tau}=17$ basis functions are sufficient to achieve a truncation error of $10^{-4}$.  We discretize the domain with 1000 finite element mesh nodes, and the spectral representation thus leads to a substantial computational savings.

We initialize the optimization procedure from a zero mean and run BFGS with the reduced deterministic objective for 50 iterations, which is sufficient to find the MAP point for the mean of both parameters.  We then perform natural gradient descent with the stochastic VFE for 2000 iterations (a characteristic trace of the cost function is shown in Fig.~\ref{fig:convergence}.  Note that because the MAP point has already been found via BFGS, the likelihood loss is already close to minimal.  The cost reduction in this optimization primarily comes from inflating the posterior variance such that it is as close to the prior as possible while still inducing model solutions that are consistent with observations.  

Fig.~\ref{fig:ismip_b_posterior}a shows samples from the posterior predictive distribution plotted alongside the noisy pseudo-observations for an array of bed undulation frequencies.  It is evident that the inference method is effective in the sense that the resulting predictive distribution is consistent with observations, but there are also differences between the samples, reflecting the fact that there are many possible solutions that are equivalently good when dealing with noisy observations.  Figs.~\ref{fig:ismip_b_posterior}b and c show the marginal posterior distributions over $\alpha$ and $\tau$ (i.e. the diagonals of the covariance matrix, which can be efficiently computed without forming the full matrix).  We see that variational inference reconstructs distribution over the traction coefficient that allow for the (constant) true value with high probability over all length scales.  However, there is substantial variance.  This is expected, as the basal velocity is, in this experiment, near zero and only a small fraction of the surface velocity; even large changes in the traction coefficient have little relative effect on the surface velocity.  Unsurprisingly, the rheologic prefactor is much more strongly constrained.  It is interesting that in the low-frequency case, the thicker and thus faster regions are very well-constrained, whereas the slow and thin regions are not well constrained by observations.  As the frequency of bedrock undulations increases, the smearing effect of substantial longitudinal stresses and a smaller signal to noise ratio lead to a higher and more homogeneous marginal variance in rheologic prefactor.

Fig.~\ref{fig:ismip_b_posterior} also shows the 3$\sigma$ marginal credibility interval of the traction coefficient when the rheologic prefactor is known exactly.  We find a substantial reduction across all length scales, with no clear spatial pattern.  In particular, we find that this additional knowledge keeps the posterior distribution away from higher traction solutions that might have been compensated for by a lower viscosity.  

\begin{figure}
    \centering
    \includegraphics{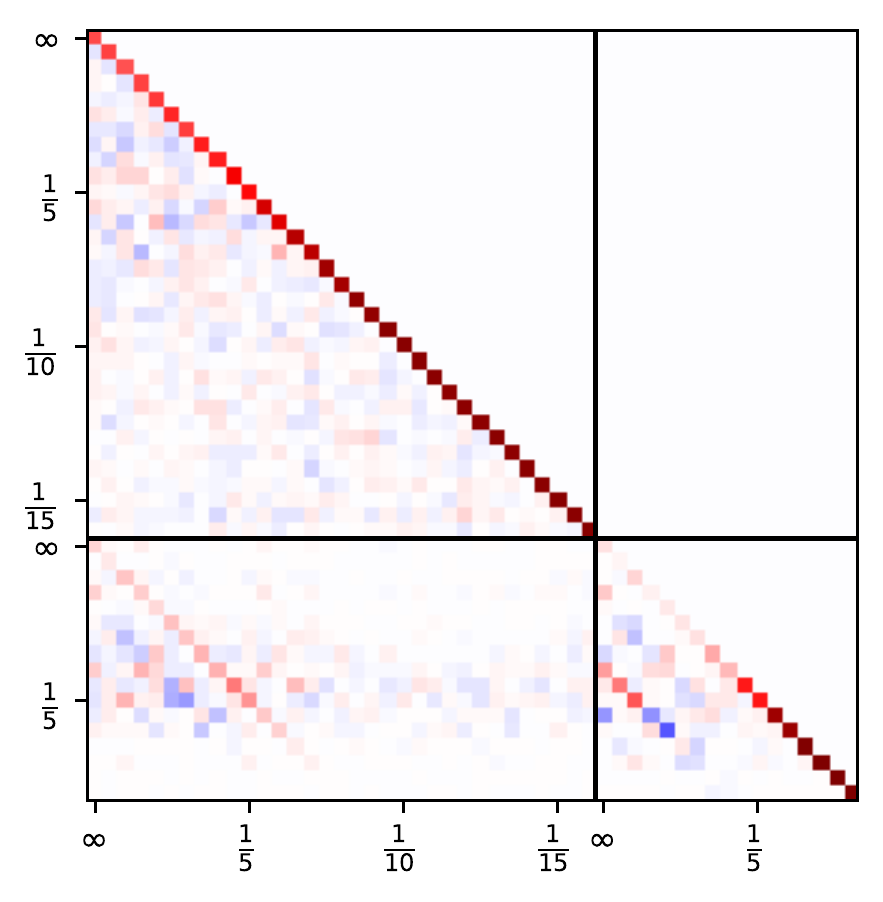}
    \caption{The unique Cholesky factor of the posterior covariance $S$ for ISMIP-HOM B, $f=2$.  White values represent zero values, while red values are positive and blues negative.  Axes are labeled as a function of spatial wavelength, with $\infty$ being constant.  The upper left block corresponds to the traction coefficient, while the lower right is softness, and the lower left their covariance.  In both cases, long wavelengths are better informed by the data than are short wavelengths, and there is substantial (and complex) correlation between the two parameters.}
    \label{fig:L_b}
\end{figure}

The direct parameterization of parameter fields via their spectral representation allows us to observe directly which length scales can be informed by the data.  Fig.~\ref{fig:L_b} shows the Cholesky factor $L$ of the posterior covariance matrix $L L^T = S$, using $f=2$ as an example.  The upper left quadrant corresponds to the modes of $\alpha$.  As expected, the diagonal entries (which roughly correspond to the standard deviation of the associated frequency) increase from top to bottom, indicating that lower frequencies are more strongly constrained by the data, while higher frequencies are relatively uninformed.  The off-diagonal entries in this block show a complex correlation structure which is a result of the spatial variability in the recovered fields.  Similarly, the lower right quadrant corresponds to the modes of $\tau$, which has a similar but even more pronounced qualitative pattern of well-informed low frequencies.  The lower left quadrant shows the covariance between modes of $\alpha$ and $\tau$; the fact that there are substantially non-zero entries implies that these two factors compensate for one another.  On the margins, this compensation makes inferences about both more uncertain.  We note that the correlation between these two variables is generally positive.  Increasing $A$ makes the ice go faster while increasing $\beta^2$ makes the ice go slower, and so it is admissible for them to vary together while yielding similar velocities.  

\subsection{ISMIP-HOM D}
\begin{figure}
    \centering
    \includegraphics{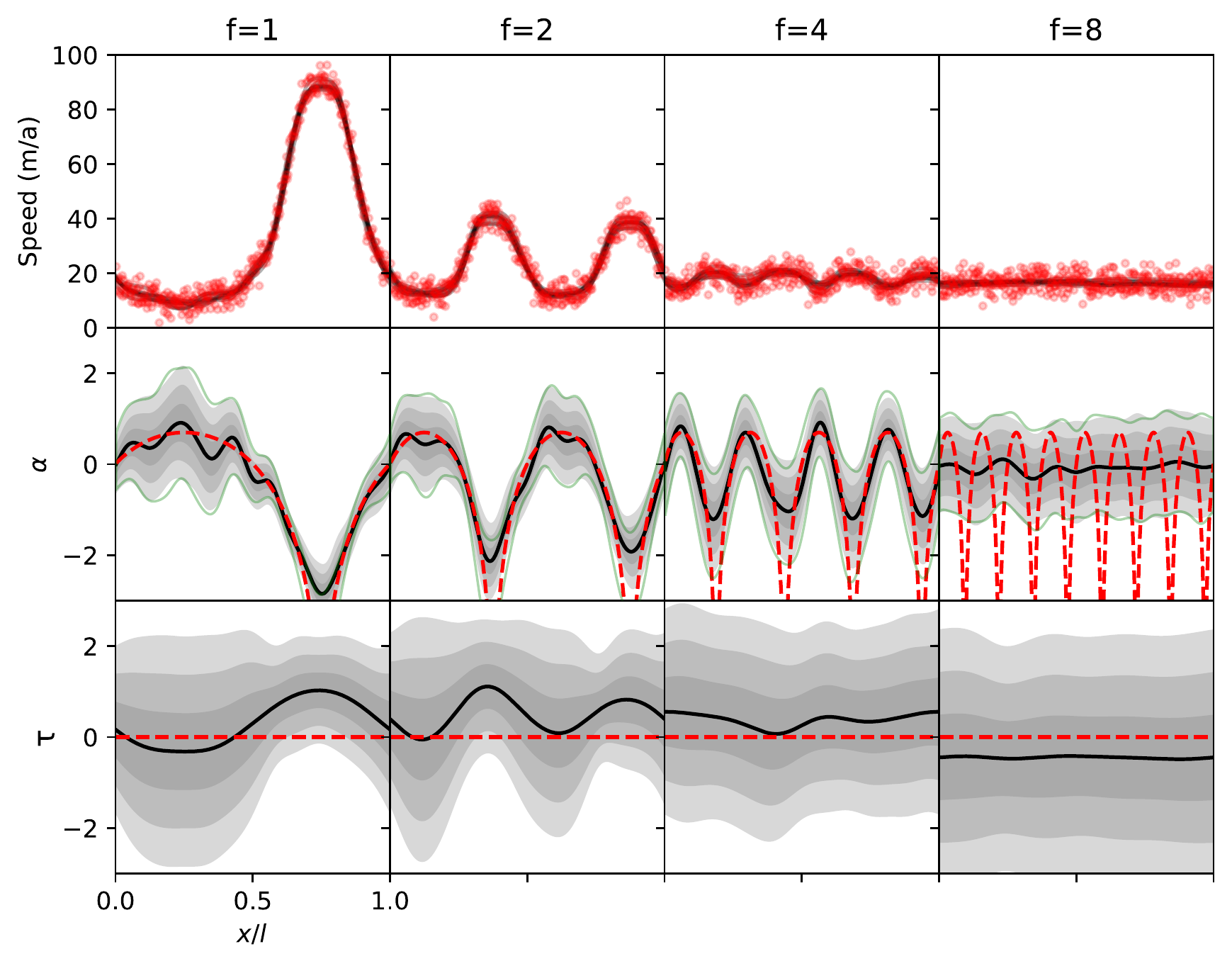}
    \caption{The posterior distribution for $f\in\{1,2,4,8\}$ for ISMIP-HOM D.  In the top row, red dots indicate observed velocity values, while overlain black lines are random samples from the posterior predictive distribution over velocity.  In the bottom two rows, the dashed red line represents the `true' parameter value, while the gray envelopes represent the 1$\sigma$, 2$\sigma$, and 3$\sigma$ marginal credibility intervals, with the black line the posterior mean.  The green lines represent the 3$\sigma$ marginal credibility interval when $\tau$ is assumed known. }
    \label{fig:ismip_d_posterior}
\end{figure}
Our ISMIP-HOM D experiment is defined over the same domain as ISMIP-HOM B, but has a few key geometric differences.  The surface elevation is given by 
\begin{equation}
    z_s = -x \tan \alpha,
\end{equation}
where $\alpha=0.1^{\circ}$ (and thus a much lower slope), and the bed elevation is
\begin{equation}
    z_b = z_s - 1000.
\end{equation}
The traction coefficient is given by 
\begin{equation}
    \beta^2 = 1001 - 1000\sin\left(\frac{2\pi f \mathbf{x}}{L}\right).
\end{equation}
Because the characteristic scale of traction coefficient is smaller in this experiment, we choose $\beta_0=10^3$.  All other aspects of the experiment are identical to those described in the previous section.

As before, Fig.~\ref{fig:ismip_d_posterior}a shows predictions plotted with pseudo-observations.  The model again shows considerable skill in matching observations, while also producing a diversity of predictions consistent with the noisy data.  Conversely to ISMIP-HOM B, the traction coefficient is well constrained.  For the low-frequency simulations, the method is able to recover the spatial pattern of traction with both high precision and accuracy.  However, we note that due to the constraints of the prior distribution and the logarithmic parameterization of this field, the method is unable to recover the traction coefficient in regions where it is near zero.  Interestingly, the inference compensates for this by increasing the ice softness rather than push the traction coefficient far from its prior.  As the frequency of traction undulations increases this effect is suppressed due to the longitudinal stresses filtering the spatial variability in surface velocities.  In this case, the most probable solution is simply to take a constant (though relatively uncertain) value for both traction coefficient and rheologic prefactor.  Again, we have also overlain the posterior distribution assuming fixed rheologic prefactor.  Interestingly, this leads to almost no difference compared to the chase where rheology is not known: the only substantive difference is in the fast flowing regions of the low-frequency experiments where the posterior variance of the fixed experiment is smaller and with a mean closer to the true value.  This lack of sensitivity is a result of two factors.  First, the glacier has very low surface slope and deformation, which is strongly dependent on the rheologic prefactor is nearly zero over all \emph{a priori} feasible prefactor values.  Second, in the low frequency cases longitudinal stresses are small, while in the high frequency cases the traction becomes effectively constant, which leads to near-zero stress divergence.  As a result, there is little information to be gained about the rhelogic prefactor from surface velocity for this experiment.  On the other hand, this lack of knowledge does not inflate uncertainty estimates in the traction coefficient.    

\begin{figure}
    \centering
    \includegraphics{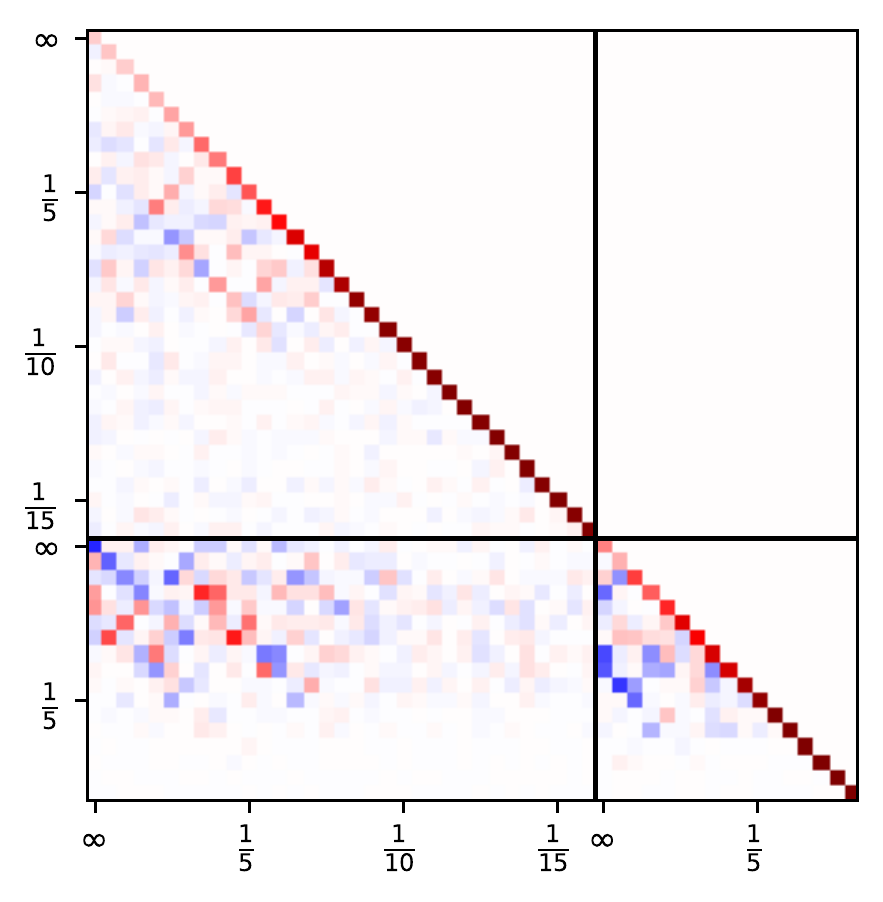}
    \caption{The unique Cholesky factor of the posterior covariance $S$ for ISMIP-HOM D, $f=2$.  White values represent zero values, while red values are positive and blues negative.  Axes are labeled as a function of spatial wavelength, with $\infty$ being constant.  The upper left block corresponds to the traction coefficient, while the lower right is softness, and the lower left their covariance.  The diagonal of the covariance between the two parameters is inverted relative to ISMIP-HOM B.}
    \label{fig:L_d}
\end{figure}

Fig.~\ref{fig:L_d} shows the Cholesky factor of the posterior covariance for $f=2$.  The story is largely the same as above in the sense that lower frequencies are better constrained by the data.  However, this time $\alpha$ is better constrained in its low frequencies than is $\tau$.  Curiously, the cross-correlations between equivalent frequencies of $\alpha$ and $\tau$ are this time negative, the opposite of the above.  However, there are also many more positive cross-correlations between non-like frequencies.  This substantially more complex cross-correlation pattern is once again most likely the result of real spatial variability in the traction coefficient.  

\section{Helheim Glacier}
\begin{figure}
    \centering
    \includegraphics{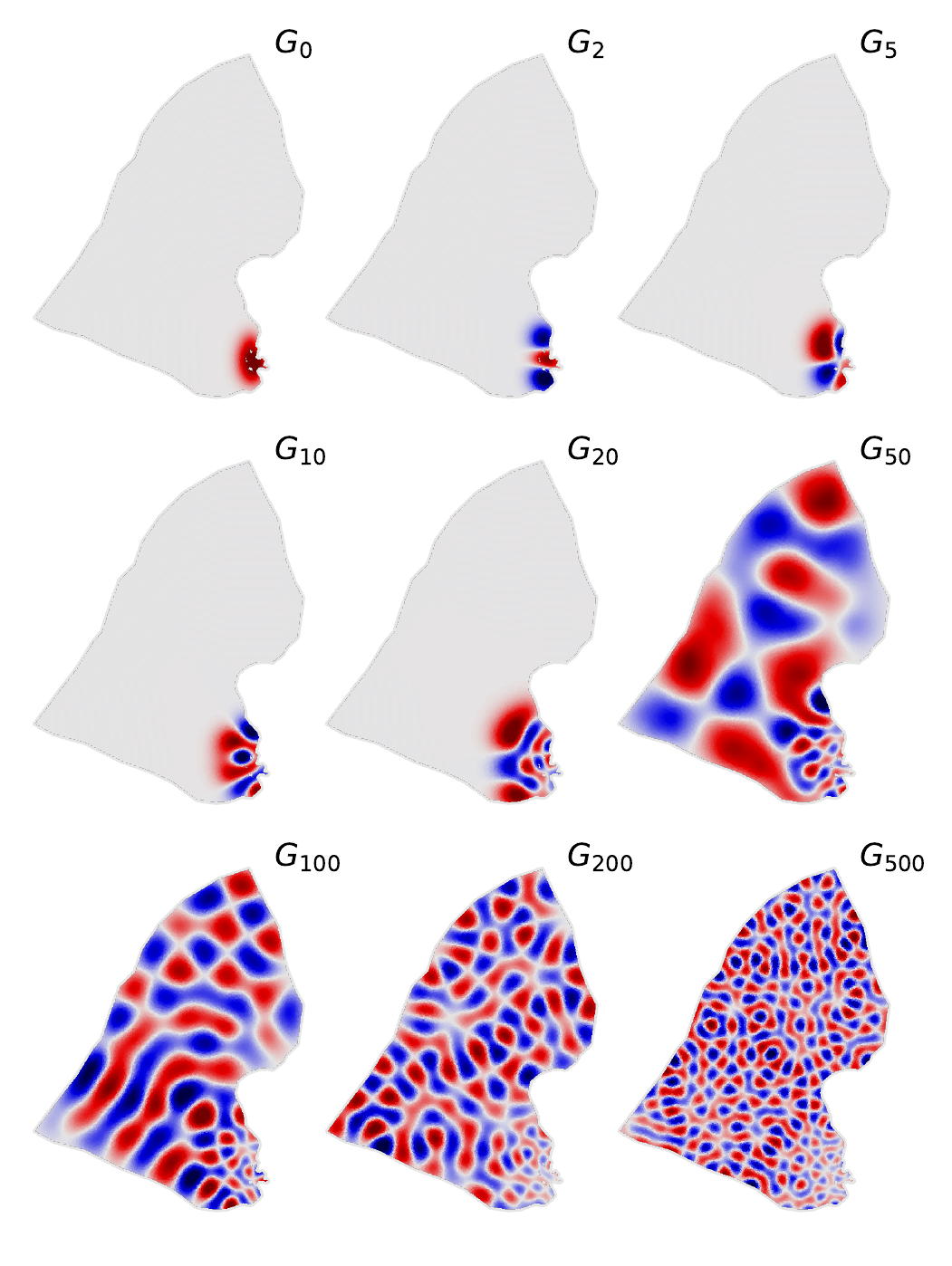}
    \caption{Select approximate eigenfunctions of the prior covariance matrix, ordered by spectral density such that lower values tend to be more important, and with red (blue) values positive (negative).  Early eigenfunctions preferentially capture low-frequency variability over the high-resolution outlet region, while later eigenfunctions capture increasingly high-frequency variations across the interior.}
    \label{fig:helheim_eigenfunctions}
\end{figure}

\begin{figure}
    \centering
    \includegraphics{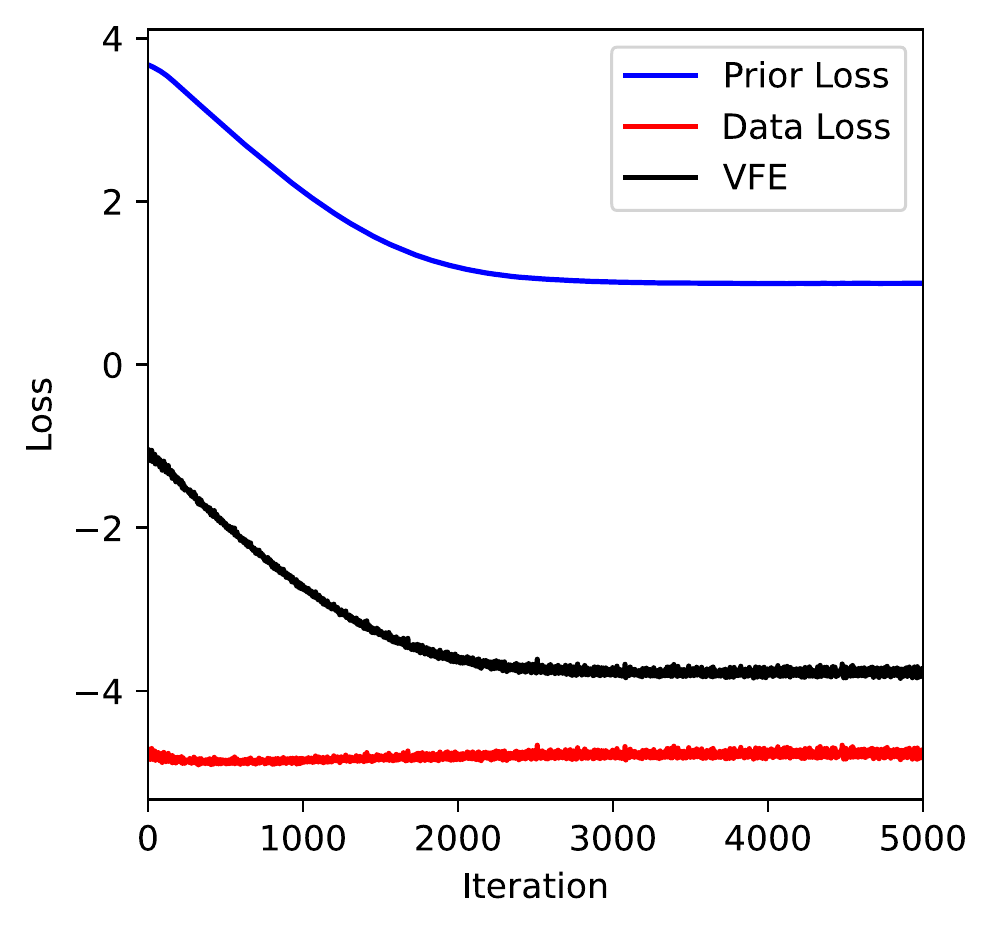}
    \caption{Components of the variational free energy as a function of natural gradient descent iteration for the Helheim experiment, indicating convergence to VFE minimum.}
    \label{fig:helheim_loss}
\end{figure}

\begin{figure}
    \centering
    \includegraphics{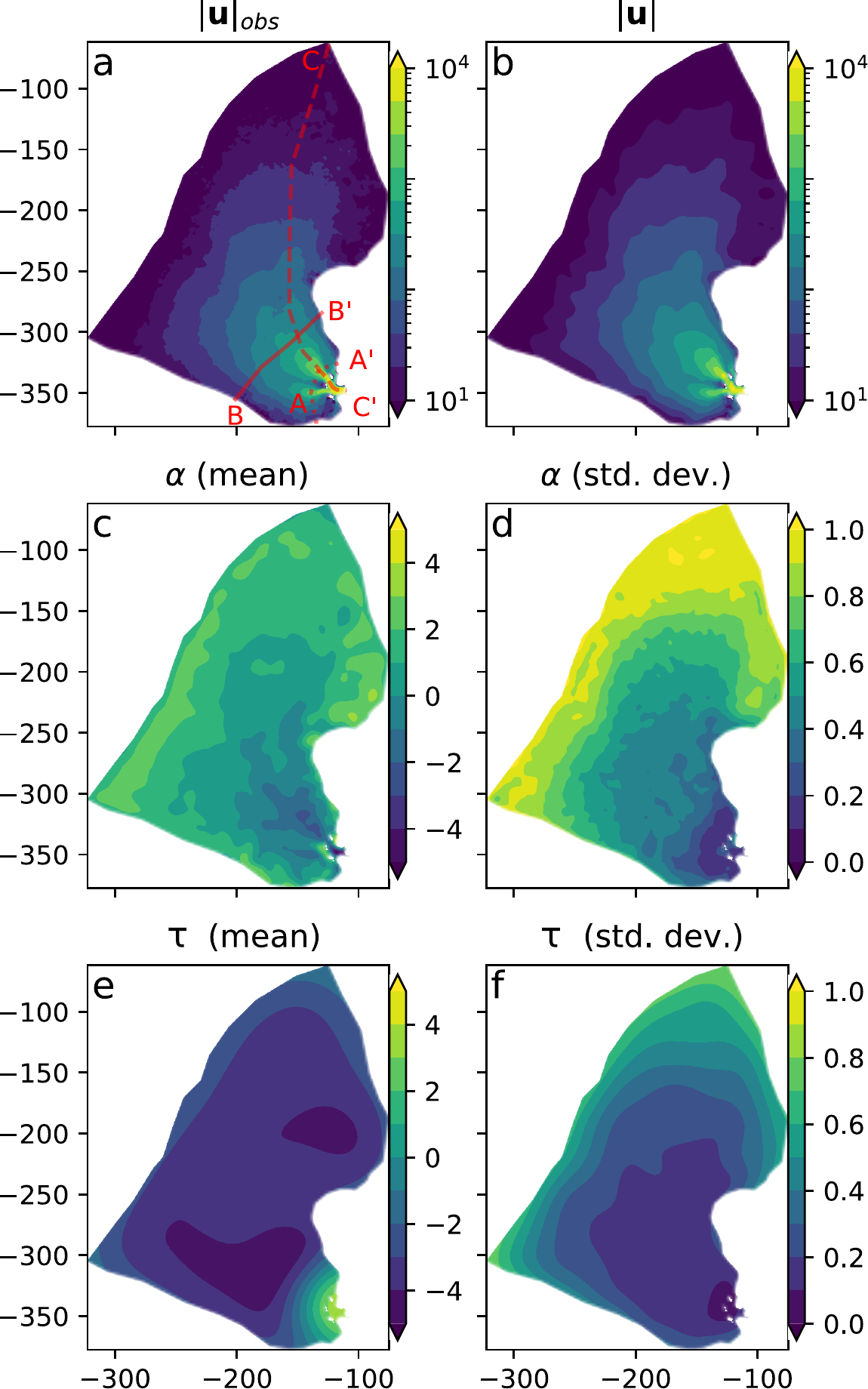}
    \caption{a) Observed surface velocity at Helheim Glacier, as well as transects illustrated in Fig.~\ref{fig:helheim_A}.  b) Mean modelled surface speed.  c) Mean of traction coefficient $\alpha$ showing a sticky interior and slippery terminus.  d) Standard deviation of traction coefficient showing marginal uncertainty in slow flowing regions that is close to the prior (which had unit standard deviation), and very low uncertainty in fast flowing regions.  e) Mean of softness parameter $\tau$, showing hard ice in the slow interior and softer ice near the terminus.  f) Standard deviation of softness parameter, again with relatively high uncertainty in the interior and low uncertainty near the terminus.} 
    \label{fig:helheim_posterior}
\end{figure}

We now turn to inferring the joint distribution over a real glacier given observations of surface velocity.  As a target we choose Helheim Glacier in southeastern Greenland.  Helheim Glacier is one of the largest outlets of the Greenland ice sheet, collecting copious moisture from the Atlantic Ocean and transporting it back to the sea through a vigorous and well-instrumented calving terminus at the head of Sermilik Fjord \citep{howat2005rapid,andresen2012rapid}.  In the past two decades, Helheim has undergone a substantial adjustment in terminus position and near-terminus flow speed.  Mean annual velocity within its basin varies spatially over four orders of magnitude, with a mean annual average speed at the glacier terminus of nearly $10^4$ma$^{-1}$.  Helheim Glacier also exhibits strong seasonal variability in terminus position and speed near the terminus.  However, in this study we neglect the latter effect and instead attempt to recover the posterior distribution over time-averaged rheologic prefactor and coefficient fields given annually averaged surface speeds.  

We specify the horizontal extent of the Helheim catchment using an ice mask packaged with the BedMachineV4 dataset \citep{morlighem2017bedmachine,morlighem2021bedmachine} to determine the position of the marine and terrestrial terminus, and by manually tracing the interior boundaries of the catchment with the aid of satellite-derived flow directions, in turn derived from MEASURES InSAR derived annual ice velocity mosaics \citep{joughin2010greenland,joughin2021measures}.  These ice velocities were also used as the observational data for the inversion.  We use thickness and surface elevation products from BedMachineV4, which uses a combination of interpolation and mass conservation principles to produce a seamless map.  We used gmsh \citep{geuzaine2009gmsh} to produce a computational mesh that varies from 400m at the mesh boundaries to 1500m in the interior, for a total of $\sim 2\times10^4$ nodal points.  

We specify the likelihood function by modelling the ice speed as independent and log-normal with $\sigma^2=0.1$, which corresponds to an uncertainty of approximately 10\%.  This is somewhat higher than the uncertainty reported with the velocity product, however this term must also alias additional sources of uncertainty, of which errors due to model uncertainty and time-averaging are the most important examples.  As a prior on traction coefficient, we specify $\beta_0=10^3$ with a length scale of $l=5$km and $\sigma^2_\alpha=1.0$.  As a prior on rheologic prefactor, we specify $A_0=10^{-17}$ with a length scale of $l=30$km and $\sigma^2_{\tau} = 1$.  We assume a data density of $\rho=\frac{1}{25}$ observations km$^{-2}$.  We find a low rank decomposition of the kernel matrix as described above to generate the feature map.  An example of these eigenfunctions for $\alpha$ are shown in Fig.~\ref{fig:helheim_eigenfunctions}.  With a truncation error of $10^{-4}$, $\alpha$ was parameterized with 1977 eigenfunctions and $\tau$ with 91.  

We use BFGS to find the MAP point, then perform stochastic natural gradient descent for 5000 iterations, with a step size of $\eta=0.1$ increasing to $\eta=3$ over the first 500 iterations.  Fig.~\ref{fig:helheim_loss} shows both prior and data loss as a function of iteration, indicating convergence to a minimum at around 2500 iterations.  As before, BFGS produces a solution for the MAP that is close to the optimal variational mean, and most of the effort of stochastic VI is in inflating the covariance matrix.  This is reflected by data loss, which becomes more noisy compared to the early stages of optimization as the covariance in $\mathbf{z}$ becomes larger.  Nonetheless, the converged variational mean is slightly higher than the MAP produced by BFGS, which is likely due to mild skewness in the true posterior distribution, perhaps due to the log-normal likelihood function.  

Fig.~\ref{fig:helheim_posterior} shows the observed and modelled surface speed means.  We note that the data has some high frequency noise that the model is not able to reproduce due to a combination of the smoothness constraints imposed by the prior on the traction coefficient (and also perhaps the smoothing effect of the ice physics model compared to noisy observations).  The figure also shows the mean and standard deviations of the model parameters.  The traction coefficient is high in interior regions where there is little sliding and low in regions of fast flow.  In particular, the main trunk (the bed of which is below sea level) has near zero traction, which is unsurprising given its marine influence .  The rheologic prefactor indicates stiff ($A<10^{-18}$) ice in much of the interior, with soft conditions near the terminus, although it remains difficult to ascribe this softening to temperature, damage, enhancement, or any other specific mechanism.  We find that uncertainty in both parameters is inversely correlated with glacier speed.  This would not be surprising given a likelihood model with a fixed uncertainty as the signal to noise ratio is lower in faster ice.  The log-normal distribution over velocities means that uncertainty scales with speed, yet faster regions are still more well-informed by the data.  In particular, we find that near the ice margins, both sliding and deformation are scantly less uncertain than under the prior.  

\begin{figure}
    \centering
    \includegraphics[width=0.8\linewidth]{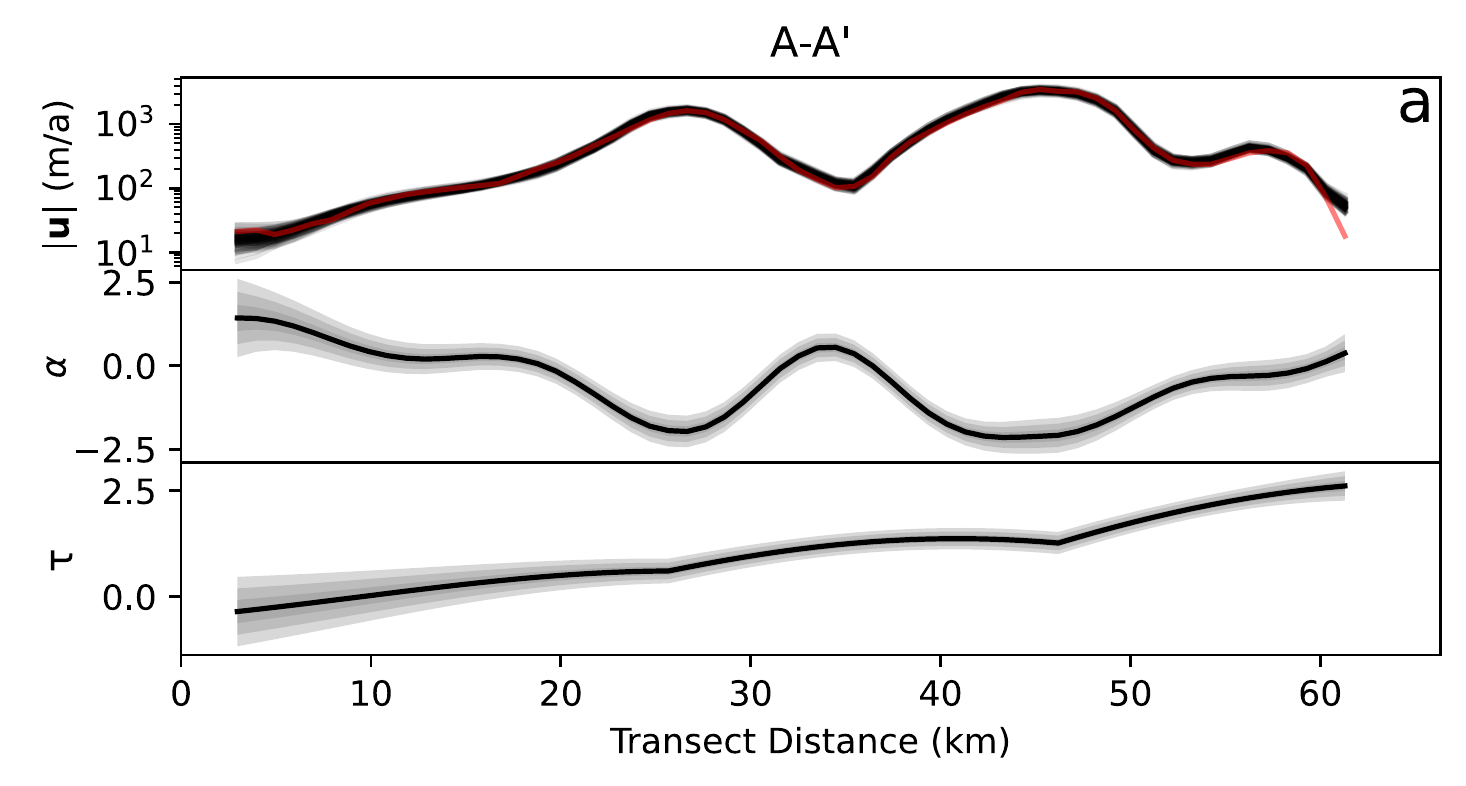}
    \includegraphics[width=0.8\linewidth]{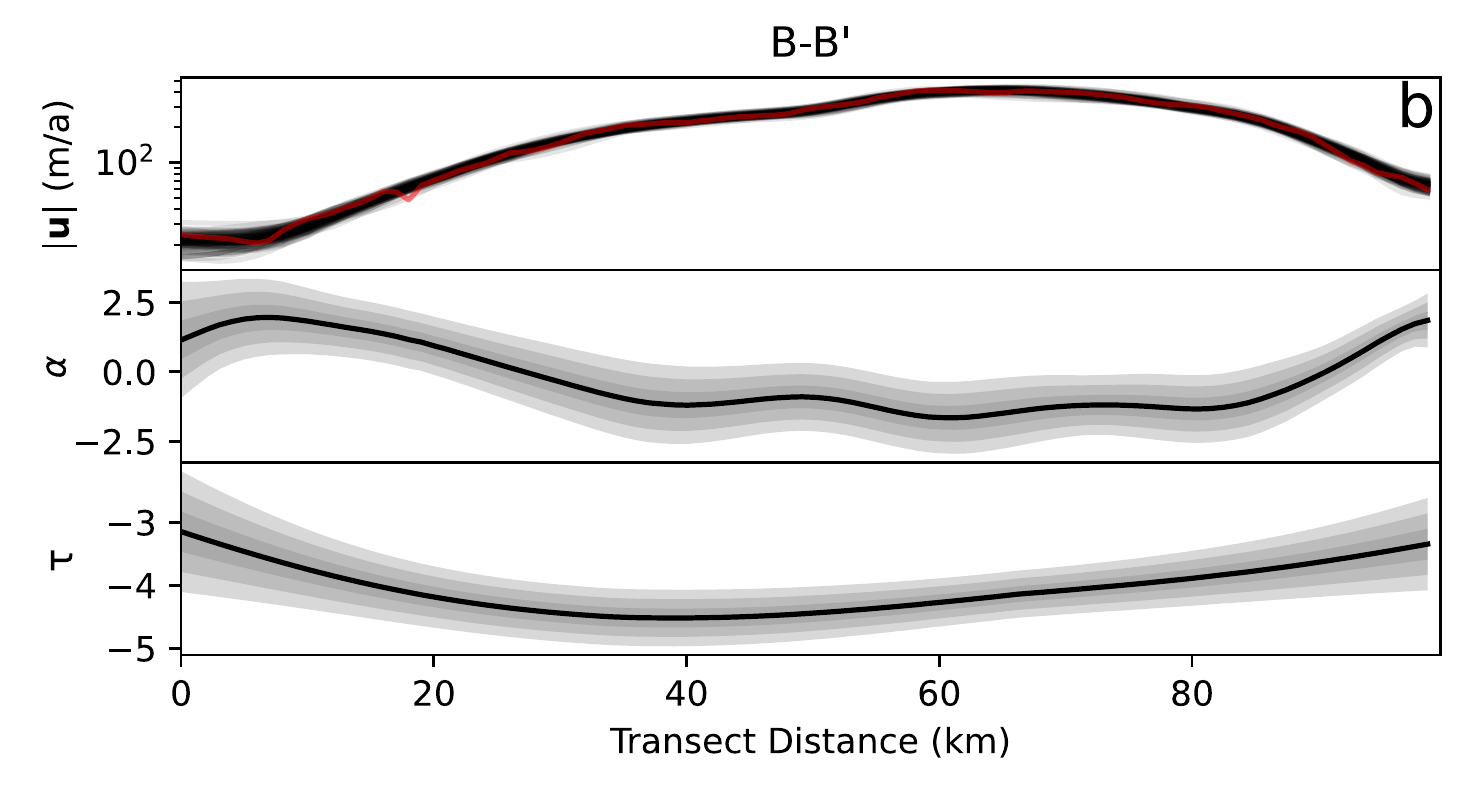}
    \includegraphics[width=0.8\linewidth]{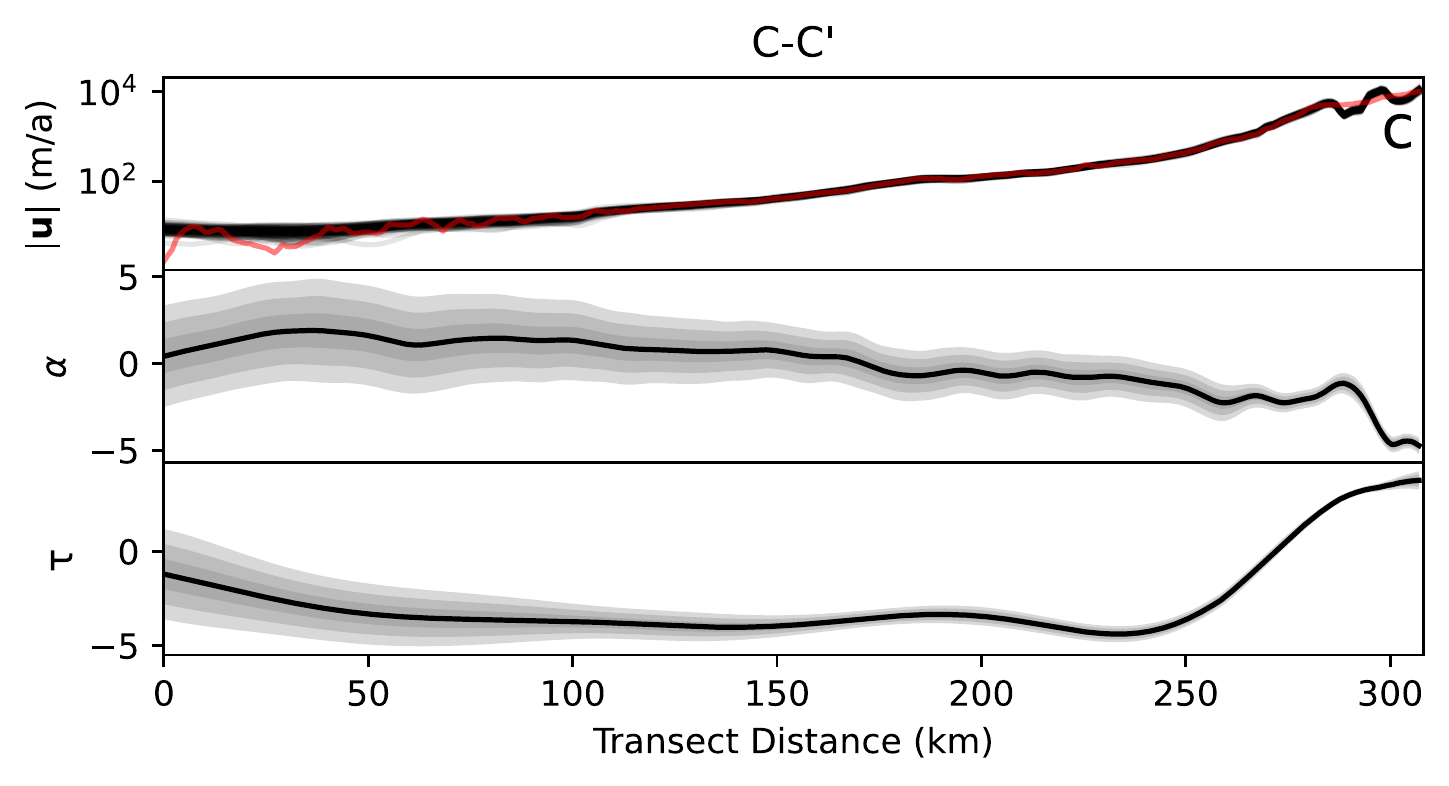}
    \caption{Posterior distributions over transects shown in Fig.~\ref{fig:helheim_posterior}, with A-A' and B-B' being cross sections, and C-C' a long profile.  Top columns show observed speed in red and samples from the posterior predictive distribution in black.  The middle and bottom rows show the distribution over the traction coefficient and rheologic prefactor, with shaded envelopes corresponding to the 1$\sigma$, 2$\sigma$, and 3$\sigma$ credibility intervals.}  
    \label{fig:helheim_A}
\end{figure}

It is also instructive to take a more detailed look at the transects indicated in Figure~\ref{fig:helheim_posterior}.  For a transect very near the terminus (Fig.~\ref{fig:helheim_A}a) we see higher uncertainties near the glacier margins.  Interestingly, the fastest regions do not correspond to the lowest uncertainties in traction coefficient.  Indeed, inflection points in the velocity (i.e. local maxima and minima in the velocity cross section) tend to have greater uncertainty than regions with greater shear (we have seen this effect in other glaciers that we have tested this method on as well).  We believe that this is due to map plane shear stress gradients (recall that these terms in the Stokes' equation are proportional to the curvature of the velocity) balancing a relatively larger fraction of the driving stress, making the velocity less sensitive to the the basal shear stress.  In a simpler and slower flowing transect (Fig.~\ref{fig:helheim_A}b), we see a much higher uncertainty in both parameters.  In this relatively thick ice, there is substantial covariance between traction coefficient and rheologic prefactor and it is not clear from surface velocity how to partition the ice flow in these regions.  An along-flow profile (Fig.~\ref{fig:helheim_A}c) shows the distinction between the slow interior and fast outlet more clearly, with uncertainty in both parameters decreasing as one traverses from interior to terminus.  

\section{Related Work}
This work builds upon substantial previous work in glaciological inverse problems.  \citet{macayeal1993tutorial} introduced the notion of using adjoint methods to find maximum likelihood (or MAP in the presence of regularization) estimates of traction coefficient with respect to observations of surface speed.  Such methods (and alternative numerical methods for solving the same problem) have now become ubiquitous as a means of model initialization and to ensure good agreement between modeled and observed surface velocities, which is a critical condition for producing correct estimates of e.g. ice flux into the ocean or a detailed analysis of the balance of stresses in glaciological systems \citep[e.g][]{joughin2004basal,morlighem2013inversion,sergienko2013regular,habermann2013changing,arthern2015flow,riel2021data}.  Similar inversion methods have also been extended to simultaneously inferring traction and auxilliary variables.  For example, \citet{gudmundsson2008limit} evaluate the degree to which both traction and bed geometry could be simultaneously inferred from surface properties.  Similar ideas are extended by \citet{perego2014optimal}, who invert for estimates of traction coefficient and bed topography so as to match velocities and also to yield a transient-free initial ice sheet state. \citet{goldberg2013parameter} and \citet{larour2014inferred} solve time-dependent problems, inverting for both ice sheet geometry and traction coefficient from observations of surface elevation and velocity.  \citet{petrat2012inexact} simultaneously invert for rheologic prefactor and traction coefficient (as we do here) using an inexact Gauss-Newton scheme, and \citet{gudmondsson2019} perform a similar inversion across the whole of Antarctica.  \citet{ranganathan2021new} again simultaneously invert for the traction coefficient and rheologic prefactor over ice streams in Antarctica, developing a specialized regularization approach in hopes of finding a unique partitioning between parameters.       

One critical limitation of point estimate approaches is that uncertainty estimates are not immediately forthcoming, which is especially important when non-uniqueness is expected due to unidentifiability between parameters.  Several methods have been proposed to characterize the Bayesian posterior distribution to various degrees of approximation; our method is closely aligned with these and we take the opportunity to emphasize where we have built upon these previous works.  \citet{petra2014computational} develop both a formalized framework for considering Bayesian inference over continuous random variables and also a Markov Chain Monte Carlo method that takes advantage of the availability of adjoint-based gradients.  Then apply this method to a set of synthetic ice sheets to generate samples from the posterior distribution of the traction coefficient.  Notably, they use the full Stokes equations (which are quite expensive to solve) for the map between parameters and observable quantities.  While such MCMC methods are highly desirable, they also require more computational power than is practical for very large problems, which is why we choose to use variational inference as an alternative.  

\citet{isaac2015scalable}, recognizing the difficulties in MCMC sampling at large scale, find the MAP solution for the traction coefficient, and then leverage the Laplace approximation, which approximates the posterior distribution as a multivariate Gaussian with mean given by the MAP and covariance given by the local curvature of the log-posterior, and efficient Hessian-vector products to find a low-rank eigendecomposition of the covariance matrix.  They apply this method at scale to Antarctica with millions of nodal points, again utilizing a full Stokes model to map from traction coefficient to velocity.  The method presented in our paper is similar in spirit, but with some substantive differences.  First, we employ a variational rather than Laplace approximation.  While both approximations are multivariate normal, the variational approximation has the advantage in that it finds the Gaussian that is closest in KL-divergence, which formally measures the difference in information content between two distributions.  From a practical perspective, what this means is that our method is sensitive to non-normality and higher moments of the true posterior (e.g. skewness and kurtosis), which is potentially important in the presence of non-linear forward models and non-Gaussian models of observational noise.  Second, we specify our prior distribution in terms of kernels, which have a rather-well defined meaning with respect to the intuitive quantities of length scale, smoothness, and prior variance.  We see this as an advantage over previous works, which instead specify the prior in terms of a single elliptic operator (either a Laplacian or biharmonic operator) for which the associated kernel and hyperparameters are not so clearly defined.  Third, while both works invoke a low-rank approximation to the posterior covariance, our representation is precomputed based on the prior covariance, whereas \citet{isaac2015scalable} compute this approximation \emph{a posteriori}.  Because our decomposition is not `tuned' to observations, we certainly do not enjoy the same scalability in the sense that we would require more basis functions to be able to capture the same posterior (assuming that the inverse posterior Hessian is a good approximation to the posterior covariance).  However, the data independence of our basis may also potentially allow for efficient estimation of prior hyperparameters (an advantage) \citep{solin2020hilbert}, although we defer explicit exploration of such to future work.  Fourth, we utilize the first-order equations (rather than full Stokes) to map from parameters to velocities, however the ramifications of such a difference are likely small \citep{cornford2020}.

In contrast to the previous works, we also construct the joint posterior over both the traction coefficient and rheologic prefactor.  \citet{babaniyi2021inferring} tackle a similar problem to ours, utilizing similar methods to \citep{petra2014computational} for quantifying the posterior covariance over a synthetic glacier.  However, they end up with a slightly different outcome: rather than the joint distribution, they find the posterior distribution over the traction coefficient alone, with additional variance resulting from marginalizing over an uncertain rheology.  In contrast, \citet{gopalan2021bayesian} uses Gibbs sampling to sample from the the joint posterior over the rheologic prefactor and traction coefficient at Langjokull, Iceland, with the former modelled as a constant and the latter as a Gaussian process (as we do here), using the shallow ice approximation as a parameter to observable map.  This latter choice decouples observation points and leads to a substantial computational savings, but may not be realistic in regions of substantial topography or fast flow.  

As a final note, a few applications of variational inference as a means to infer spatially and temporally variable parameters with likelihood functions defined by partial differential equations have appeared in problems outside of glaciology.  \citet{barajas2019approximate} demonstrate the effective recovery of the PDF of diffusivity in a 1D diffusion problem, as well as the distribution over prior hyperparameters, while \citet{yang2017} solve for a complex 2D diffusivity field on the unit square.  \citet{franck2016sparse} adopt a similar approach as ours (including a dimensionality-reducing decomposition of a complex parameter field) for a complex problem in elastography.  Finally, \citet{yang2021b} have applied variational inference to perform Bayesian inference over physics-informed neural networks (i.e. a spectral method for solving PDEs using neural networks as adaptive basis functions), recovering a diffusivity field for both a 1D non-linear diffusion model and a 2D diffusion-reaction model.

\section{Discussion}
\subsection{Subjectivity of the the prior and possible avenues for improvement}
The choice of prior strongly influences the solution of the inversion, yet was chosen in an \emph{ad hoc} fashion, both with respect to the functional form and also with respect to the specific hyperparameter values.  This is a significant limitation.  However, one potential advantage of the variational inference framework is that the negative VFE is a lower bound on the marginal likelihood, i.e. the likelihood of observing the data integrated over all possible values of traction coefficient and rheologic prefactor.  Thus by minimizing the VFE with respect to model hyperparameters (perhaps simultaneously with the parameters themselves), it is possible to adaptively tune prior hyperparameters \citep{salimbeni2018natural}.  This can be thought of as conceptually similar to the ubiquitous L-curve analysis, in which regularization parameters are increased until an inflection point in the fit to observations is found \citep{calvetti2000}.  We hope to explore this possibility in future work.  With respect to the form of the prior (e.g. a Gaussian process with a manually specified covariance function), it is not clear that this PDF is a sensible one.  \citet{arthern2015exploring} demonstrate a particularly interesting avenue for future inquiry by deriving a prior for the traction coefficient that is as ignorant in the sense that it maximizes entropy subject to the physical constraints of the problem.  While it is not clear that such methods are practically applicable for a problem such as this one (where considerable numerical effort is needed simply to perform inference with a specified prior), the goal of finding a more objective prior is an important one that needs to be considered to produce more credible applications of Bayesian inference.  

\subsection{Scaling to larger systems}
Our ability to quantify the full posterior covariance matrices is contingent upon our ability to represent the parameters of interest as coefficients of a more parsimonious basis.  We have shown that this is plausible at the scale of individual (large) glaciers with the spectral basis that we developed here, reducing the degrees of freedom for Helheim glacier from $\sim$20,000 nodes to $\sim$2000 eigenfunctions.  We note that the vast majority of these eigenfunctions were associated with the quickly varying traction, rather than the slowly varying rheology, because the number of significant basis functions scales with the ratio of the domain size to the prior length scale, exponentiated to the dimensionality of the problem.  As such, a low rank representation at the scale of the entire Greenland ice sheet would involve an order of magnitude more basis functions.  Since our inference algorithm scales with $m^3$, such a representation strains at the edges of computational feasibility.  As such, the spectral representation presented here is not suitable for larger systems (in the sense of geography, not in terms of degrees of freedom of the underlying mesh, which our method is independent of outside of the computation of the forward and adjoint model).  This is a critical limitation that needs to be addressed.  

A few potential alternatives exist.  The simplest is the classic mean field approximation, which would model the parameters directly as uncorrelated nodal values.  Unfortunately, because covariances are explicitly ignored, marginal variance tends to be underestimated.  There exists a middle ground in the form of sparse covariance matrices, such that the covariance matrix is only non-zero for locations within a certain length scale of one another.  Similarly, a diagonal plus low rank approximation to the covariance matrix may be feasible, although the computation of natural gradients is not easy in either case.  One particularly encouraging possibility is in the use of an inducing point formulation where the parameters of interest would be defined over limited set of pseudo-inputs with learnable coordinates, allowing for adaptive resolution with a fixed computational cost \citep[e.g.][]{titsias2009variational}.  Another possibility would be the use of hierarchical matrices, which use interleaved full- and low-rank blocks to effect $\mathcal{O}(n\log^2 n)$ complexity for computing matrix roots (one of the expensive bottleneck operations required for stochastic VI) \citep{ambikasaran2015fast,petra2019hierarchical}.  We hope to explore these alternatives in future work.    

\subsection{Mass conservation}
While variational inference in general has potential for applications in a wide range of glaciological settings, one obvious application to the methods developed here is in the creation of so-called `mass-conserving beds,' or the inference of thickness fields that simultaneously satisfy whatever observations of velocity and thickness exist, as well as the mass conservation equation
\begin{equation}
    \nabla \cdot \bar{\mathbf{u}} H = \dot{a} - \frac{\partial H}{\partial t},
    \label{eq:mass_conservation}
\end{equation}
where $\dot{a}$ and $\frac{\partial H}{\partial t}$ are (perhaps imperfect observations of) the specific mass balance and thickness rate of change.  \citet{brinkerhoff2016bayesian} addressed this problem using GP priors and MCMC in the case of a 2D glacier centerline; however, it was not clear how such a method might be scaled to larger and higher dimensional problems.  The methods presented here prescribe a way forward, with the Stokes' equations replaced with Eq.~\ref{eq:mass_conservation}.  Better still, the physical constraints presented in this work could be combined with Eq.~\ref{eq:mass_conservation}, the parameter vector augmented with distributed ice thickness, and the set of observations with observations of sparse ice thickness from, for example, radar soundings.  The joint posterior distribution over such a distribution would be useful for defining initial conditions for ensembles of ice sheet models, and would represent the complete distribution of states consistent with both physical and observational constraints. 

\section{Conclusion}
We have presented a new method for Bayesian inference of the joint distribution of basal traction coefficient and rheological parameters in glaciers and ice sheets, given observations of surface velocity and an ice sheet model that maps parameters to observations.  Rendering this problem computationally tractable required a few tricks.  First, we reduced the dimensionality of the problem by casting it in terms of the eigenbasis of the prior covariance.  While direct decomposition of the prior covariance is computationally intractable, we skirted this issue by using a reduced-rank approximation based on Hilbert space methods.  Second, we used natural gradient descent to define a more appropriate metric for distance between probability distributions, and used this method to minimize the Kullback-Leibler divergence between the true posterior distribution and a candidate variational distribution, which we assumed to be a Gaussian process with the same form as the prior distribution.  

Having developed a methodology, we then applied it to two of the ISMIP-HOM experiments.  We found that capturing the covariance structure between rheologic prefactor and traction coefficient leads to a substantial inflation of posterior covariance, as either a perturbed traction coefficient or perturbed rheologic prefactor could induce an equivalent perturbation in surface speeds.  Nonetheless, we find that our method was able to effectively characterize the complete array of traction coefficient and rheologic prefactor settings that were both consistent with assumed smoothness and magnitude constraints (as encoded by a covariance function) and the observations.  

Finally, we inferred the posterior joint distribution at Helheim Glacier in Southeast Greenland.  We found substantial uncertainty in both fields, particularly in slow flowing regions, with much more informed parameter fields in the fast flowing outlet glaciers.  For Helheim glacier, we found that the mean traction field largely mirrors the surface velocity, with high traction coefficients in the interior and low traction coefficients in the outlet.  Simultaneously, we found relatively stiff ice in the interior with very soft ice in the immediate vicinity of the glacier outlet, although we cannot speculate as to whether this implies ice that is temperate, damaged, or softer for a different reason.  

The method presented here represents the first application of Bayesian inference to multiple parameters at scale and also the first application of variational inference.  Despite presenting applications to inferring the traction coefficient and rheologic prefactor, we believe that the methods here are applicable to a broad range of PDE-constrained inference problems encountered in glaciology.  

\section*{Acknowledgements}
This work was funded through Heising-Simons Foundation Grant 2019-1157.  A Jupyter notebook that performs the ISMIP-HOM experiments and a series of scripts for reproducing the Helheim Glacier results are available at github.com/douglas-brinkerhoff/vi\_at\_glacier\_scale.

\begin{appendices}
\section{Approximation of the feature map with Hilbert space methods}
The Weiner-Khintchin theorem establishes a Fourier duality between a kernel and its spectral density 
\begin{align}
    k(\mathbf{r}) = \frac{1}{2\pi^d} \int S(\omega) \mathrm{e}^{\mathrm{i} \omega^T \mathbf{r}} \mathrm{d}\omega \label{eq:kernel-spectral-density}\\
    S(\omega) = \int k(\mathbf{r}) \mathrm{e}^{\mathrm{i} \omega^T \mathbf{r}} \mathrm{d} \mathbf{r}, \label{eq:spectral-density}
\end{align}
where $\mathbf{r} = \mathbf{x} - \mathbf{x}'$, $\omega$ is the frequency, and $S(\omega)$ the spectral density, which provides the contribution of a given frequency to a GP with kernel $k(\mathbf{r})$, i.e. for a smooth kernel with long length scale, $S(\omega)$ goes to zero quickly for increasing $\omega$, while rougher and shorter length scale kernels admit higher frequency variability.  Critically, if the kernel is isotropic and stationary, then the spectral density depends only on the norm of $\omega$.  The squared exponential covariance function has spectral density
\begin{equation}
    S_{SE}(\omega) = \sigma^2 (2\pi l^2)^{d/2}\mathrm{exp}\left(-\frac{1}{2}\omega^2\ell^2\right),
\end{equation}
which is both isotropic and stationary.

Expanding the spectral density in powers of $|\omega|^2$, we have
\begin{equation}
S(|\omega|) = \sum_j a_j |\omega|^j.
\label{eq:density}
\end{equation}
The inverse Fourier transform of the spectral density defines the integral operator
\begin{equation}
    \mathcal{K} f(\mathbf{x}) = \int k(\mathbf{x},\mathbf{x}') f(\mathbf{x}') \mathrm{d}\mathbf{x}',
    \label{eq:kernel_operator}
\end{equation}
which is simply the convolution of the kernel over a function.  Using the identity $\mathcal{F}[\nabla^2 f](\omega) = -|\omega^2|\mathcal{F}[f](\omega)$ and taking the inverse Fourier transform of Eq.~\ref{eq:density}, we get
\begin{equation}
\mathcal{K} = \sum_j a_j (-\nabla^2)^j,
\end{equation}
which has the interesting implication that the covariance operator associated with a kernel may be written as a weighted sum of powers of the Laplacian.  However, still need to identify the coefficients $a_j$  

Consider now the eigenfunctions $\phi_j(\mathbf{x})$ and eigenvalues $\lambda_j$ over a domain $\Psi$ such that
\begin{align}
    -\nabla^2 \phi_j(\mathbf{x}) &= \lambda_j \phi_j(\mathbf{x}), \; \mathbf{x} \in \Psi\\
    \phi_j(\mathbf{x}) &= 0, \; \mathbf{x} \in \partial \Psi \label{eq:eigenboundary}
\end{align}
where $\Psi$ is a compact domain and $\partial \Omega$ its boundary.  Mercer's theorem states that we can write a kernel for the Laplacian as 
\begin{equation}
    \hat{l}(\mathbf{x},\mathbf{x}') = \sum_j \lambda_j \phi_j(\mathbf{x}) \phi_j(\mathbf{x}'),
\end{equation}
such that
\begin{equation}
    -\nabla^2 f(\mathbf{x}) = \int_\Psi \hat{l}(\mathbf{x},\mathbf{x}') f(\mathbf{x'}) \mathrm{d}\mathbf{x}',
\end{equation}
subject to boundary conditions Eq.~\ref{eq:eigenboundary}.  Because the eigenfunctions $\phi_j(\mathbf{x})$ are orthonormal, we can take powers of the Laplacian as
\begin{equation}
    (-\nabla^2)^s f(\mathbf{x}) = \int_\Psi \hat{l}^s(\mathbf{x},\mathbf{x}') f(\mathbf{x'}) \mathrm{d}\mathbf{x}', \label{eq:laplacian_power}
\end{equation}
with 
\begin{equation}
    \hat{l}^s(\mathbf{x},\mathbf{x}') = \sum_j \lambda_j^s \phi_j(\mathbf{x}) \phi_j(\mathbf{x}').
\end{equation}
If we take a linear combination of both sides of Eq.~\ref{eq:laplacian_power}, we get
\begin{equation}
\sum_k a_k (-\nabla^2)^k f(\mathbf{x}) = \sum_k \int a_k \hat{l}^k(\mathbf{x},\mathbf{x}') f(\mathbf{x}') \mathrm{d} \mathbf{x}.
\end{equation}
The left hand side is the covariance operator $\mathcal{K}$ applied to $f(\mathbf{x})$.  Substituting  Eq.~\ref{eq:kernel_operator} and integrating over $\Psi$, we have
\begin{equation}
    \int_\Psi k(\mathbf{x},\mathbf{x}') f(\mathbf{x}') \mathrm{d}\mathbf{x}' = \int_\Psi \sum_k a_k \hat{l}^k(\mathbf{x},\mathbf{x}') f(\mathbf{x}') \mathrm{d} \mathbf{x},
\end{equation}
and thus we can immediately identify the kernel
\begin{align}
    k(\mathbf{x},\mathbf{x}') &= \sum_k a_k \hat{l}^k(\mathbf{x},\mathbf{x}')\\
                              &= \sum_j \sum_k a_k \lambda_j^k \phi_j(\mathbf{x}) \phi_j(\mathbf{x}').
\end{align}
Finally, letting $|\omega|^2=\lambda_j$, and substituting Eq.~\ref{eq:spectral-density}, we have that
\begin{equation}
    k(\mathbf{x},\mathbf{x}') = \sum_j S(\sqrt{\lambda_j}) \phi_j(\mathbf{x}) \phi_j(\mathbf{x}'). 
    \label{eq:eigenkernel}
\end{equation}
Eq.~\ref{eq:eigenkernel} states that any kernel can be represented as a weighted sum over the eigenfunctions of the Laplacian, subject to boundary conditions, with weights given by the spectral density.  

For simple domains, the Laplacian eigensystem has an analytical solution.  For example, defined over a 2D rectangle $\Psi = [0,\chi_1] \times [0,\chi_2]$, the eigenfunctions of the Laplacian subject to Dirichlet boundary conditions are
\begin{equation}
    \phi_{jk} = \sqrt{\frac{2}{\chi_1 \chi_2}} \sin \left(\pi j \frac{\mathbf{x}_1}{\chi_1}\right) \sin\left(\pi k \frac{\mathbf{x}_2}{\chi_2}\right),
    \label{eq:rectangle_eigenfunctions}
\end{equation}
with eigenvalues
\begin{equation}
    \lambda_{jk} = \left(\frac{\pi j}{\chi_1}\right)^2 + \left(\frac{\pi k}{\chi_2}\right)^2. 
    \label{eq:rectangle_eigenvalues}
\end{equation}
It is simple to define $\chi_1$ and $\chi_2$ and rescale coordinates such that this rectangle encompasses the index set $\mathbf{X}$, extending far enough beyond the bounds of $\mathbf{X}$ such that the effects due to boundary conditions are negligible.  If we truncate the sum in Eq.~\ref{eq:eigenkernel} at some $M<<n$, and letting the indices $j,k$ in Eqs.~\ref{eq:rectangle_eigenfunctions} and \ref{eq:rectangle_eigenvalues} be flattened into a single sum over $M$, we can write down a feature map
\begin{equation}
    Q = \begin{bmatrix} \phi_{1}(\mathbf{X} & \ldots \phi_{M}(\mathbf{X}) \end{bmatrix} \mathrm{diag} \left(\begin{bmatrix} \sqrt{S(\sqrt{\lambda_{1}})} \\ \vdots \\ \sqrt{S(\sqrt{\lambda_{M}})} \end{bmatrix} \right),
\end{equation}
where we have dropped $\diamond$-notation, but note that this can be done independently for independent GP priors.  

We can exact additional computational savings by specializing this feature map to the index set.  Noting that $K(\mathbf{x},\mathbf{x}') \approx Q Q^T$, we can efficiently find its eigenvalues which are shared with $Q^T Q$ (which is in $\mathbb{R}^{M \times M}$).  Solving this smaller eigenvalue problem is $\mathcal{O}(M^3)$ but needs only be done once.  We can then recover the eigenvectors scaled by sqrt of these eigenvalues by using the identity 
\begin{equation}
    G = \tilde{V} \mathrm{diag}(\sqrt(\lambda)) = Q V,
\end{equation}
where $\tilde {V}$ contains the eigenvectors of $Q Q^T$ and $V$ the eigenvectors of $Q^T Q$ (the careful reader will note that this could also be done via singular value decomposition; however the SVD would be $\mathcal{O}(nM^2)$, whereas we only solve a $\mathcal{O}(M^3)$ Hermitian eigenproblem.  Constructing the matrix $Q^T Q$ and evaluating $Q V$ are still $\mathcal{O}(nM^2)$, but these have a substantially smaller leading constant.  As such, the asymptotic complexity is equivalent but doing it this way leads to a significant practical savings).  We can again truncate the decomposition with $\lambda$ below a threshold, which will typically lead to a substantial reduction in retained features.  Interestingly, experimentation on small problems has shown that the specialization described above leads to feature maps with columns that are identical to those computed by direct eigendecomposition of the kernel matrix, and we can view this method as a general mechanism for constructing the rank $m$ eigendecomposition of an arbitrary kernel in $\mathcal{O}(nm^2)$ time, where we assume that $M=\mathcal{O}(m)$. 

\section{Ice Sheet Model}
We compute the map from traction coefficient and rheologic prefactor to observable surface velocities $\mathbf{u}(\mathbf{x};\beta^2,A)$ as the solution to a set of discretized partial differential equations described here.
  The flow of the ice sheet over a volume $\Omega$ is modelled as a low Reynolds number fluid using a hydrostatic approximation to Stokes' equations \citep{pattyn2003new}
\begin{align}
    \nabla \cdot \tau' = \rho_i g \nabla z_s,
    \label{eq:stokes}
\end{align}
where 
\begin{align}
    \tau' = \begin{bmatrix} 2\tau_{xx} + \tau_{yy} & \tau_{xy} & \tau_{xz} \\
                            \tau_{xy} & \tau_{xx} + 2\tau_{yy} & \tau_{yz}
    \end{bmatrix}.
\end{align}
$z_S$ is the glacier surface elevation, $\rho_i$ is ice density, $g$ the gravitational acceleration, and $\tau_{ij}$ is a component of the deviatoric stress tensor given by
\begin{equation}
\tau_{ij} = 2\eta \dot{\epsilon}_{ij},
\end{equation}
with $\dot{\epsilon}$ the symmetrized strain rate tensor.  The viscosity 
\begin{equation}
\eta = \frac{A(\mathbf{x})}{2}^{-\frac{1}{n}}(\dot{\epsilon}_{II} + \dot{\epsilon}_0)^{1-\frac{1}{n}}
\end{equation}
is dependent on the second invariant of the strain rate tensor $\dot{\epsilon}_{II}$.  The exponent in Glen's flow law $n=3$.

At the ice surface $\Gamma_{z_s}$ and terrestrial margin $\Gamma_{T}$ (where the ice thickness is assumed to approximate zero), we impose a no-stress boundary condition
\begin{align}
    \tau' \cdot \mathbf{n} = \mathbf{0},
\end{align}
where $\mathbf{n}$ is the outward pointing normal vector, and $\mathbf{0}$ is the zero vector.  At marine boundaries $\Gamma_M$, we impose a normal pressure condition
\begin{align}
    \tau' \cdot \mathbf{m} = P_w \mathbf{n},
\end{align}
where $P_w$ is the water pressure, typically computed as $\rho_w g d$, with $d$ the depth below sea level.  

The remaining lateral boundary $\Gamma_L$ is synthetic in the sense that there are no natural physical boundary conditions that should be applied there.  At these boundaries we apply the 'no boundary' condition \citep{griffiths1997no}, in which we maintain the boundary term arising from integration by parts of a weak form, and which is equivalent to setting a stress free condition at an infinitely distant boundary.  

Finally, at the basal boundary $\Gamma_{z_B}$ we impose the linear sliding law
\begin{equation}
\label{eq:sliding_law}
\tau'\cdot \mathbf{n} = -\beta^2 \mathbf{u}.
\end{equation}

We discretize the momentum equations using a mixed finite element method.  Introducing a terrain-following $\varsigma$-coordinate
\begin{equation}
\varsigma = \frac{z_s-z}{H},
\end{equation}
where $z_s$ is the upper ice surface, $H$ is ice thickness and $z$ the vertical coordinate, we decompose the domain as $\Omega = \bar{\Omega} \times [0,1]$.  Introducing a test function $\Psi(x,y,\varsigma)$, multiplying it by Eq.~\ref{eq:stokes}, and integrating over the domain, we obtain the following variational formulation: find $\mathbf{u} \in \Upsilon$, such that
\begin{align}
    \label{eq:disc_stokes}
    0 &= \int_{\bar{\Omega}} \int_1^0 (\bar{\nabla} \Psi + \partial_\varsigma \Psi \bar{\nabla} \varsigma) \cdot \tau' \,H \;\mathrm{d}\varsigma \;\mathrm{d}\Omega \nonumber \\
      & - \int_{\Gamma_l} \int_1^0 \Psi \cdot \tau' \cdot \mathbf{n} \; \mathrm{d} \varsigma \mathrm{d}\Gamma - \int_{\Gamma_M} \int_1^0 \Psi \cdot \mathbf{n} P_w \nonumber \\
      & - \int_{\bar{\Omega}} \int_1^0 \Psi \cdot \tau_d \;\mathrm{d}\varsigma\;\mathrm{d}\Omega \nonumber \\
      & + \int_{\bar{\Omega}} \Psi \cdot \beta^2 \mathbf{u} \,\mathrm{d}\Omega |_{\varsigma=1}, \nonumber \\
      & \forall\Psi\in V,
\end{align}
where $\bar{\nabla}$ is the gradient operator in the two map-plane dimensions and $\tau_d=\rho g H \bar{\nabla} z_s$ is the gravitational driving stress, and with $\Upsilon\in W^{1,2}(\Omega)$, and where $W^{1,2}$ is a Sobolev space over the model domain $\Omega$.  To discretize the weak form, we utilize the Galerkin approximation and restrict $\Psi$ to a finite subset of $\Upsilon$:
\begin{equation}
    \Psi \in \hat{\Upsilon} \subset \Upsilon,
\end{equation}
where 
\begin{equation}
\hat{\Upsilon} = \Upsilon_{\bar{\Omega}} \otimes \Upsilon_{\bar{\Omega}} \otimes \Upsilon_0 \otimes \Upsilon_0
\end{equation}
is a tensor product of function spaces defined over $\bar{\Omega}$ and $[0,1]$, respectively.  For $\Upsilon_{\bar{\Omega}}$, we use the continuous piecewise linear Lagrange basis $\left\{\upsilon_i\right\}_{i=1}^{n_p}$, where $n_p$ is the number of grid points in a mesh defined on $\bar{\Omega}$ \citep{zienkiewicz2005finite}.  For $\Upsilon_0$, we utilize the basis set 
\begin{equation}
    \left\{ \psi_1 = 1, \psi_2 = \frac{1}{n+1} [(n+2) \varsigma^{n+1} - 1] \right\}.
\end{equation}
We introduce the ansatz solution
\begin{equation}
\mathbf{u}(x,y,\varsigma) = \sum_{i\in n}\left[\bar{\mathbf{u}}_i + \mathbf{u}_{d,i} \frac{1}{n+1} [(n+2)\varsigma^{n+1} - 1] \right]\upsilon_i(x,y),
\end{equation}
where $\bar{\mathbf{u}}$ is the vertically averaged velocity, and $\mathbf{u}_d$ is the deviation from that average induced by vertical shearing.  The above expression implies that the solution in the vertical dimension is a linear combination of a constant (i.e. the shallow-shelf approximation) and a polynomial of order $n+1$, which corresponds to the analytical solution of the isothermal shallow ice approximation. 

We use the finite element software FEniCS  \citep{logg2012automated} to compile all of the variational problems described above.  We solve the problems over an isotropic computational mesh with variable resolution, ranging from approximately 250m diameter elements near the margins to approximately 1km near the ice divide.  The mesh was created using a Delaunay Triangulation routine in the package gmsh \citep{geuzaine2009gmsh}.  We use Newton's method to find the root of the non-linear residual, using a Jacobian inferred from an automated symbolic computation of the G{\^a}teaux derivative.  Note that this implies that we must solve a large linear system of equations each Newton iteration.  We solve this system using ILU-preconditioned GMRES as implemented in PETSc \citep{balay2019petsc}.  
\end{appendices}

\bibliographystyle{apalike}
\bibliography{references}

\end{document}